\documentclass[aps,prb,reprint,twocolumn,superscriptaddress,amsmath,amssymb,amsfonts]{revtex4-1}

\AtBeginDocument{\usepackage{booktabs}}               
\makeatletter
\g@addto@macro\bfseries{\boldmath}
\makeatother

\usepackage[varg]{txfonts}
\usepackage{chngcntr}
\usepackage{color}
\usepackage{hyperref}
\definecolor{cL}{RGB}{32,145,140}
\hypersetup{colorlinks=true,linkcolor=cL,citecolor=cL,urlcolor=cL} 

\usepackage{amsmath}
\usepackage{graphicx}
\usepackage[percent]{overpic}
\usepackage{mathrsfs}
\usepackage{bm}
\usepackage{braket}
\usepackage[nolist,nohyperlinks]{acronym}
\usepackage{natbib}
\usepackage{microtype}

\newcommand{\tsub}[1]{\textsubscript{#1}}

\newcommand{\subref}[2]{\ref{#1}\hyperref[#1]{#2}}

\newcommand{\avg}[1]{\braket{#1}}

\renewcommand{\vec}[1]{\boldsymbol{#1}}

\newcommand{\abo}[2]{#1\tsub{2}#2\tsub{2}O\tsub{7}}
\newcommand{\eto}{\abo{Er}{Ti}}
\newcommand{\yto}{\abo{Yb}{Ti}}

\newcommand{\ygo}{\abo{Yb}{Ge}}

\newcommand{\meV}{\ {\rm meV}}
\newcommand{\T}{\ {\rm T}}

\newcommand{\K}{\ {\rm K}}
\newcommand{\mK}{\ {\rm mK}}

\newacro{DM}[DM]{{Dzyaloshinskii-Moriya}}

\newcommand{\ybge}{\ygo{}}

\newcommand{\ybti}{\yto{}}
\newcommand{\ybsn}{Yb$_2$Sn$_2$O$_7$}

\newcommand{\luge}{\abo{Lu}{Ge}}

\begin{document}
%
%
%
%
%
\title{Unravelling competing microscopic interactions at a phase boundary: \\ a single crystal study of the metastable antiferromagnetic pyrochlore \ybge{}}

\author{C. L. Sarkis}
\affiliation{Department of Physics, Colorado State University, 200 W. Lake St., Fort Collins, Colorado 80523-1875, USA}
\author{J. G. Rau}
\email{jrau@uwindsor.ca}
\affiliation{Max-Planck-Institut f\"ur Physik komplexer Systeme, 01187 Dresden, Germany}
\affiliation{Department of Physics, University of Windsor, Windsor, Ontario, N9B 3P4, Canada}
\author{L. D. Sanjeewa}
\affiliation{Department of Chemistry, Clemson University, Hunter Chemistry Laboratory, Clemson, South Carolina 29634,USA}
\author{M. Powell}
\affiliation{Department of Chemistry, Clemson University, Hunter Chemistry Laboratory, Clemson, South Carolina 29634,USA}
\author{J. Kolis}
\affiliation{Department of Chemistry, Clemson University, Hunter Chemistry Laboratory, Clemson, South Carolina 29634,USA}
\author{J. Marbey}
\affiliation{Department of Physics and National High Magnetic Field Laboratory, Florida State University, 1800 East Paul Dirac Drive, Tallahassee, Florida 32310, USA}
\author{S. Hill}
\affiliation{Department of Physics and National High Magnetic Field Laboratory, Florida State University, 1800 East Paul Dirac Drive, Tallahassee, Florida 32310, USA}
\author{J. A. Rodriguez-Rivera}
\affiliation{NIST Center for Neutron Research, National Institute of Standards and Technology, Gaithersburg, Maryland 20899, USA}
\affiliation{Materials Science and Engineering, University of Maryland, College Park, Maryland 20742, USA}
\author{H. S. Nair}
\affiliation{Department of Physics, University of Texas El Paso, 500 W University Ave, El Paso, Texas 79902}
\author{D.R. Yahne}
\affiliation{Department of Physics, Colorado State University, 200 W. Lake St., Fort Collins, Colorado 80523-1875, USA}
\author{S. S{\"a}ubert}
\affiliation{Department of Physics, Colorado State University, 200 W. Lake St., Fort Collins, Colorado 80523-1875, USA}
\author{M. J. P. Gingras}
\affiliation{Department of Physics and Astronomy, University of Waterloo, Waterloo, ON, N2L 3G1, Canada}
\affiliation{Quantum Materials Program, CIFAR, MaRS Centre,
West Tower 661 University Ave., Suite 505, Toronto, Ontario, M5G 1M1, Canada}
\author{K. A. Ross}
\email{Kate.Ross@colostate.edu}
\affiliation{Department of Physics, Colorado State University, 200 W. Lake St., Fort Collins, Colorado 80523-1875, USA}
\affiliation{Quantum Materials Program, CIFAR, MaRS Centre,
West Tower 661 University Ave., Suite 505, Toronto, Ontario, M5G 1M1, Canada}

\date{\today}
%
%
%
%
\begin{abstract}
We report inelastic neutron scattering measurements from our newly synthesized  single crystals of the structurally metastable antiferromagnetic pyrochlore \ybge{}. We determine the four symmetry-allowed nearest-neighbor anisotropic exchange parameters via fits to linear spin wave theory supplemented by fits of the high-temperature specific heat using the numerical linked-cluster expansion method. The exchange parameters so-determined are strongly correlated to the values determined for the $g$-tensor components,  as previously noted for the related Yb pyrochlore \ybti{}. To address this issue, we directly determined the $g$-tensor from electron paramagnetic resonance of 1\% Yb-doped \luge{}, thus enabling an unambiguous determination of the exchange parameters. Our results show that \ybge{} resides extremely close to the classical phase boundary between an  antiferromagnetic $\Gamma_5$ phase and a splayed ferromagnet phase. By juxtaposing our results with recent ones on \ybti{}, our work illustrates  that the Yb pyrochlore oxides represent ideal systems for studying quantum magnets in close proximity to classical phase boundaries. 
\end{abstract}

\maketitle


\section{Introduction}
Phase competition in correlated electron systems is intimately linked to their novel behavior, such as high $T_c$ superconductivity~\cite{grissonnanche2014direct}, colossal magnetoresistance~\cite{dagotto2005complexity}, and the formation  of quantum spin liquids (QSLs)~\cite{savary2016quantum}. Such complex systems with competing, or frustrated, interactions exhibit rich phase diagrams with many phase boundaries, as vividly
illustrated by highly-frustrated magnets (HFM)~\cite{HFM_book}.
Near phase boundaries ---- regions of strongest competition ----  quantum fluctuations can play an important role in shifting the phase boundaries, reducing the average order parameter, or potentially  producing intrinsically quantum states such as QSLs or valence bond/plaquette order~\cite{chandra1988possible,capriotti2000spontaneous, cabra2011quantum,reuther2011magnetic,gong2015quantum}. Finding materials that lie close to classical phase boundaries can thus provide invaluable insights into the effects of competing quantum many-body interactions, and result in the discovery of new phenomena.  If a material, or family of materials, is thought to exhibit this phase competition, it is essential to determine precisely the nature of the microscopic interactions. To do so, the study of high-quality single crystals is crucial, since orientational averaging from polycrystalline (powder) samples can obscure important features, such as the excitation spectra. 

In this work, we take advantage of the new availability of single crystals of the Yb pyrochlore \ybge{} to determine its microscopic exchange interactions and show that the \abo{Yb}{M} family of pyrochlore oxides are exquisite materials for studying exotic phase boundary effects in HFM systems. This knowledge should motivate future studies aimed at tuning these materials directly to the phase boundary, while also providing an important benchmark for improving our theoretical understanding of anisotropic exchange~\cite{rau2018yb} in the now widely-studied class of Yb-based quantum magnets
\cite{hester2019novel,bordelon2019field,ranjith2019field,wu2019tomonaga,ranjith2019anisotropic,sala2019crystal,rau2016anisotropic}.

\begin{figure*}[tb]
\includegraphics[width =0.75\textwidth]{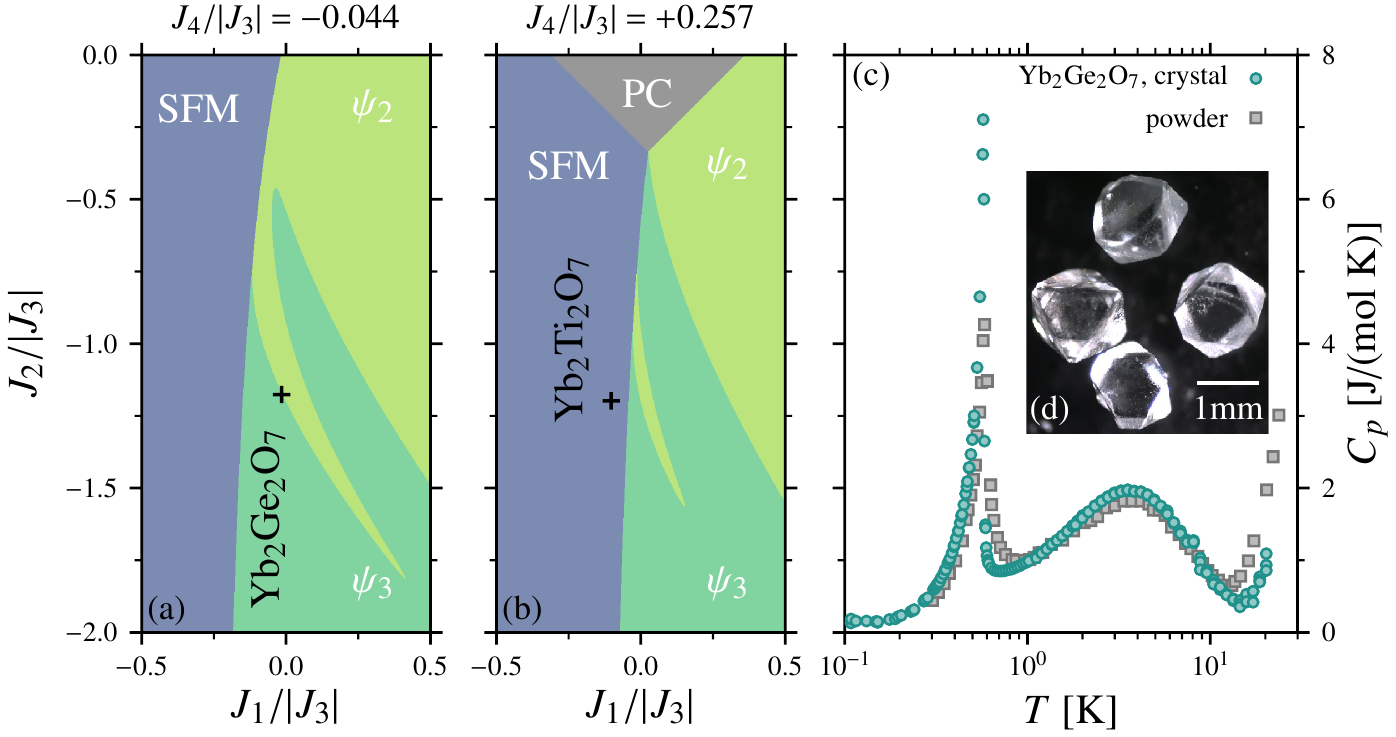}
\caption{(a,b) Sections of classical phase diagram for the anisotropic exchange model [Eq.~(\ref{eqn:hamiltonian})] relevant for (a) \ybge{} and (b) \yto{}. (c) Comparison of the specific heat of \ybge{} from a representative single crystal~\cite{sanjeewa2018single} and the powder sample studied in Ref.~\onlinecite{dun2015antiferromagnetic}. (d) Optical images of representative single crystals, adapted from Ref.~\onlinecite{sanjeewa2018single}. } 
\label{fig:phasediag}
\end{figure*}

The rare-earth pyrochlore lattice materials beautifully exemplify the diversity of behaviors possible for competing interactions in frustrated effective spin-$\frac{1}{2}$ systems~\cite{hallas2018experimental,rau2019frustrated}. At typical experimental temperature and energy scales, the angular momentum of the magnetic rare-earth ions can often be described as a pseudo-spin-$\frac{1}{2}$ with anisotropic exchange interactions~\cite{rau2019frustrated}. Detailed inelastic neutron scattering (INS) studies on single crystals of rare-earth titanate pyrochlores~\cite{ross2011quantum,savary2012order,robert2015spin,thompson2017quasiparticle} have cemented the acceptance of a unifying minimal physical model~\cite{ross2011quantum,savary2012order,zhitomirsky2012quantum} that underlies the behavior of many of these materials. This model is the nearest-neighbor (effective) spin-$\frac{1}{2}$ anisotropic exchange Hamiltonian for the pseudo-spins $\vec{S}$,
\begin{equation}
    H = \sum_{\avg{ij}} \sum_{\mu\nu} J_{ij}^{\mu\nu}S_{i}^{\mu}S_{j}^{\nu} - \mu_{\textrm{B}}\sum_{\mu\nu} B^{\mu} \sum_{i} g_{i}^{\mu\nu}S_{i}^{\nu},
    \label{eqn:hamiltonian}
\end{equation}
where $\mu$ and $\nu$ run over the Cartesian directions $(x,y,z)$, $J_{ij}^{\mu\nu}$ is the exchange tensor between spins at lattice sites $i$ and $j$, $g_{i}^{\mu\nu}$ is the $g$-tensor for spin at site $i$, and $B^\mu$ is the $\mu$ component of the external magnetic field. 

For the pyrochlore lattice, symmetry allows four independent exchange parameters ($J_{1}, J_2, J_3, J_{4}$)~\cite{curnoe2008structural,ross2011quantum}. Varying these exchanges, the \emph{classical} phase diagram contains four $\vec{q}=0$ ordered phases: three antiferromagnetic phases (the $\psi_2$, $\psi_3$ and Palmer-Chalker (PC) states) and one splayed ferromagnet phase (SFM)~\cite{WongPRB2013,yan2017theory}. The Yb pyrochlore oxides \ybti{}, \ybge{} and \ybsn{} are prime candidates for realizing strong phase competition described by this model. While \ybge{} has been found to order into a $\Gamma_5$ AFM ground state ($\psi_2$ or $\psi_3$)~\cite{dun2015antiferromagnetic}, both  \ybti{}~\cite{yasui2003ferromagnetic,gaudet2016gapless, yaouanc2016novel} and \ybsn{}~\cite{yaouanc2013dynamical,lago2014glassy} order into SFM states. This strongly suggests that these three materials lie close to a phase boundary between a $\Gamma_5$ phase and an SFM phase. To date, this has only been verified for \ybti{}~\cite{robert2015spin,thompson2017quasiparticle,rau2019magnon,scheie2019multiphase} due to the availability of large single crystals of that material.

 In order to shed light on the evolution of this Yb series through the magnetic phase diagram and to assess the proximity of  \ybge{} to a boundary
 with \emph{any} potentially competing phase(s), we have studied a collection of single crystals which were recently grown 
 hydrothermally~\cite{sanjeewa2018single}. We determined the exchange parameters for \ybge{} and found that it is (classically) as close to the SFM/$\Gamma_5$ phase boundary as \ybti{},  but now \emph{within} the $\Gamma_5$ phase, with the leading quantum fluctuations predicted to select a $\psi_3$ state.

\section{Experimental Methods}

The cubic pyrochlore structure of \ybge{} (room temperature lattice parameter $a=9.8297(7) \AA$~\cite{sanjeewa2018single}) is a metastable phase. The thermodynamically stable crystal structure is the tetragonal pyrogermanate~\cite{dem1988synthesis,becker1987phases,cai2011tolerance,sanjeewa2018single}, but the pyrochlore structure has been previously obtained as powder samples by high pressure and high temperature synthesis (1300$^\circ$C, 6 GPa)~\cite{hallas2016xy,shannon1968synthesis}. The growth of large single crystals that could readily be used for INS investigations is not yet possible under these extreme conditions, though Ho$_2$Ge$_2$O$_7$ and Dy$_2$Ge$_2$O$_7$ have been prepared very recently as small crystals under high pressure~\cite{antlauf2019synthesis}. Meanwhile, a relatively low temperature hydrothermal synthesis (650$^\circ$C) can stabilize the pyrochlore structure of \ybge{} and produce high quality single crystals of approximately $1\times1\times1$ mm$^{3}$ size~\cite{dem1988synthesis,sanjeewa2018single}. Clear and colorless single crystals of cubic \ybge{} were synthesized by this method [Inset of Fig.~\subref{fig:phasediag}{(d)}]. 

The temperature dependence of the specific heat, $C_p(T)$, measured on a 0.67 mg single crystal, was previously reported~\cite{sanjeewa2018single}; we reproduce it here for comparison to the powder data from ~\citet{dun2014chemical,dun2015antiferromagnetic} 
[Fig.~\subref{fig:phasediag}{(c)}]. A broad feature centered around 3.5 K, and a sharp peak at $T_{\textrm N} = 0.572(4)$ K, are observed. Such features have been argued to correspond to the onset of short-range spin correlations and long-range order, respectively, in Yb pyrochlores~\cite{hallas2016xy,hallas2016universal,Applegate2012a,Hayre2012}. The good agreement between the powder and the single crystal $C_p(T)$ data, the colorless appearance of the crystals, as well as  the x-ray refinement results of Ref.~[\onlinecite{sanjeewa2018single}], indicate that ``stuffing'' defects, or other non-idealities of the crystal structure that could produce a sample dependence of the physical properties~\cite{ross2012lightly, sala2014vacancy, arpino2017impact}, are negligible in our single crystals of \ybge{}. Magnetic susceptibility data, $\chi(T)$, on the same single crystal (not oriented) were obtained using a vibrating sample magnetometer from 100 K down to 1.8 K, in a field of 100 Oe.  

Continuous wave electron paramagnetic resonance (EPR) spectra were recorded from a 50 mg collection of micro-crystals~\footnote{This collection of relatively large (compared to powder samples) crystals represents a large sampling of random orientations, but does not exactly correspond to a powder average. We did not pulverize the crystals in order to avoid strain broadening of the $g$-tensor.} of 1\% Yb doped Lu$_2$Ge$_2$O$_7$ (Lu$_{1.98}$Yb$_{0.02}$Ge$_2$O$_7$) which were synthesized in a similar manner as the \ybge{} crystals~\cite{sanjeewa2018single}.  Several EPR spectra were taken at varying temperatures using a superheterodyne quasi-optical setup operating at 120 GHz, described in Ref.~[\onlinecite{van2005quasioptical}]. Data at different temperatures were taken in order to observe that the resonance peak positions in the dilute compound do not shift, thus eliminating any  possible concerns of  spin interactions affecting the determination of the $g$-tensor. 
\begin{figure*}
	\includegraphics[width = \textwidth]{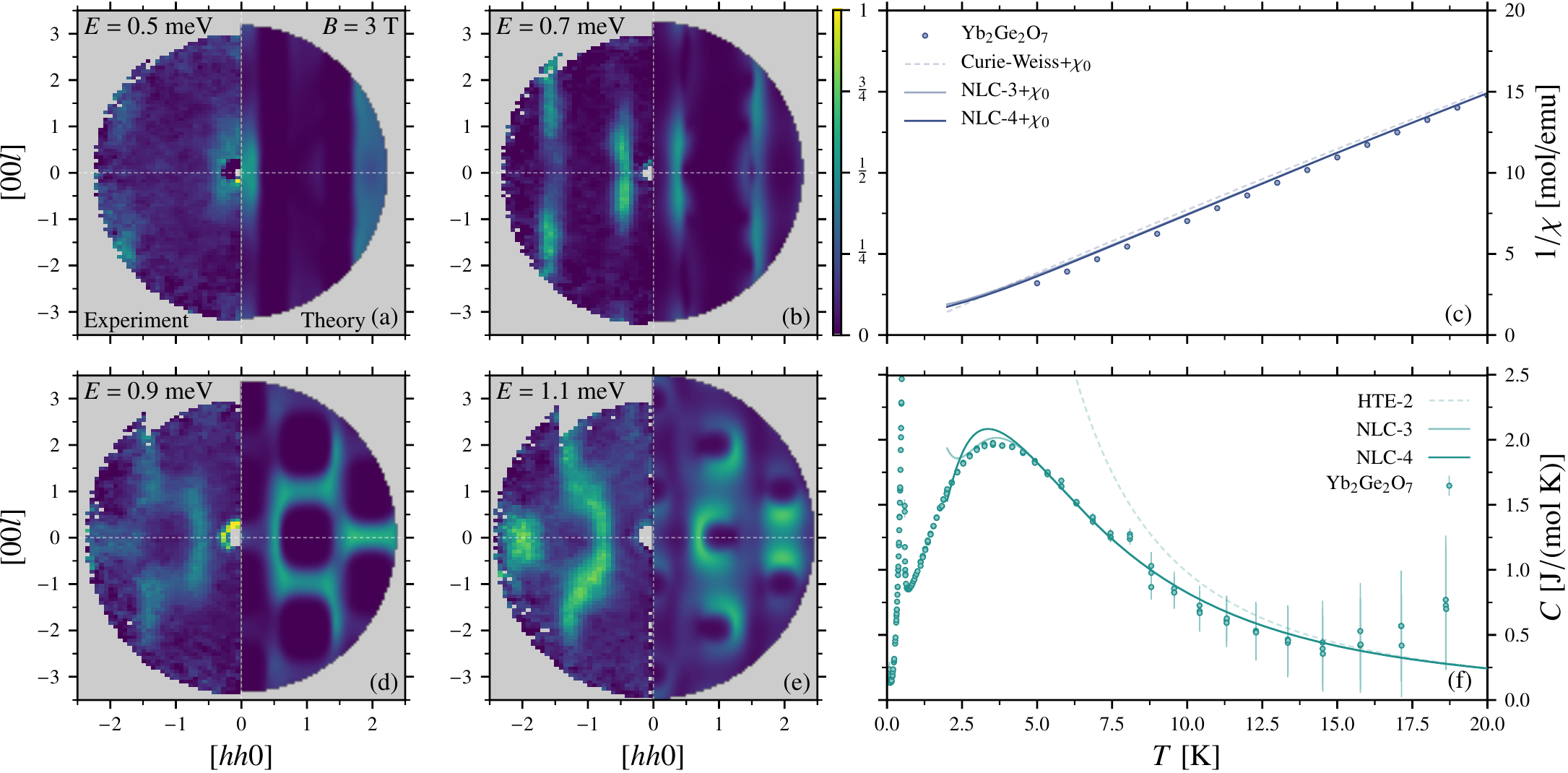}
	\caption{(a,b,d,e) Comparison of constant-energy slices (centered at energy $E$ with an energy dependent energy resolution function~\ref{sec:fitting}) of the 3 T field polarized spin-waves between \ybge{} at $1.8\K$ (left) and linear spin wave theory using the best fit $J_1$-$J_4$ parameters within Eq.~(\ref{eqn:hamiltonian}) (right). The overall intensity scale is consistent between panels, but arbitrary. Comparison between the (c) magnetic susceptibility and (f) specific heat and NLC calculations for the parameters listed in Table~\ref{tab:params}. }
	\label{fig:INS}
\end{figure*}
INS data were collected using the Multi Axis Crystal Spectrometer (MACS)~\cite{rodriguez2008macs} at the NIST Center for Neutron Scattering, under an applied field up to $9$ T, and the Cold Neutron Chopper Spectrometer (CNCS) at the Spallation Neutron Source in zero field~\cite{ehlers2011new}.  Twenty-eight single crystals of \ybge{} (total mass $\sim160$ mg) were co-aligned on an oxygen-free copper mount to orient the $[hhl]$ plane horizontally and the field vertically along the $[1\bar{1}0]$ direction.  The overall  mosaic spread of the crystal array was found to be $\leq 5^{\circ}$ (Appendix \ref{sec:INS}). At MACS, INS data were taken throughout the $[hhl]$ plane at a constant energy-transfer ($E = |E_{\rm f} - E_{\rm i}|$), using a fixed final energy of $E_{\rm f}=3.7$ meV and varying $E_{\rm i}$, in a configuration that produces an energy resolution of 0.17 meV at the elastic line.  

Although the sample was in a dilution refrigerator with base temperature of the mixing chamber reading 260 mK, comparison of our zero field base temperature data with data taken at $1.8\K$ suggests that the sample did not cool below this higher temperature (see discussion in Appendix \ref{sec:INS}).  We therefore assign a temperature of 1.8 K to our field-polarized INS measurements presented here.  This higher temperature does not affect the spin wave dispersions in the field-polarized paramagnetic state; since the excitation energies are large relative to the temperature with the effects of the relevant Bose factor being negligible.  Corresponding constant energy $[hhl]$ slices are shown in Fig.~\ref{fig:INS}, where each energy slice took approximately 5 hours. The 3 T data, with 9 T data used as a background subtraction, was used in conjunction with thermodynamic and EPR data to determine the exchange parameters, as described below. The zero-field CNCS data collection parameters are described in Appendix \ref{sec:zerofield}.

\section{Results and Discussion}

\subsection{Results}
\label{sec:results:a}
 First, we address the single-ion properties of Yb${}^{3+}$ in \ybge{}. The site symmetry of Yb in \ybge{} is trigonal ($D_{3d}$). This results in two independent $g$ factors: one in the local $xy$ plane ($g_\pm$) and one along the local $z$ direction ($g_z$). Studies of \ybti{} have shown that fitted $g$-tensor values and exchange parameters are strongly correlated when fitting field-polarized INS~\cite{thompson2017quasiparticle}. An unambiguous determination of the $g$-values, independent from the fitting of the exchange parameters, is thus essential. Guided by this lesson, we used EPR to directly measure the $g$-tensor of 1\% Yb doped \luge{} on a randomly oriented collection of micro-crystals. 
 
The measurements were performed on a coarse powder of micro-crystalline material (1\% Yb$^{3+}$ in \luge) to avoid any sample degradation that might result from over-grinding to the degree usually necessary for powder EPR studies. Because of this, many sharp, albeit weak resonances corresponding to individual randomly oriented micro-crystals can be observed in between the extremes of the spectra; these resonances give the appearance of an increased noise level, but they are real signals from individual micro-crystals. The sample was remeasured multiple times after stirring to confirm a re-distribution of the stronger signals. 

\begin{figure}
\includegraphics[width = 0.9\columnwidth]{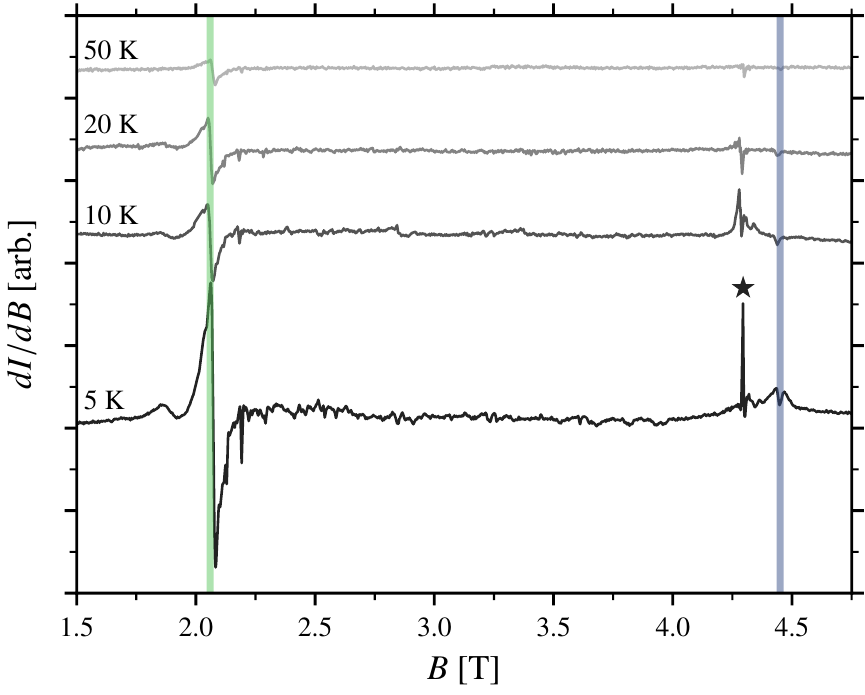}
\caption{EPR spectra from a 50mg collection of 1\% Yb-doped \luge{} microcrystallites at different temperatures (arbitrary scale, offset for clarity). The $g$-tensor values are determined from the minimum and maximum of the broad EPR absorption, $g_\pm = 4.20(5)$, $g_z = 1.93(2)$, which are shown with the green and blue vertical lines. An unidentified $g \approx 2$ impurity is present (indicated by the star).}
\label{fig:EPR} 
\end{figure}

The principal components of the $g$-tensor were determined from the end-points of the 120 GHz EPR absorption profile. These end-points manifest as a first derivative in the recorded spectrum (see Fig.~\ref{fig:EPR}) due to the use of field modulation and lock-in detection method of the in-phase signal. For an axial spectrum ($g_x$ = $g_y$ $\neq$  $g_z$), as expected on the basis of the local site symmetry at the Yb sites, sharp features in the first derivative spectrum are expected only at the onset and cessation of the absorption profile, i.e., the end-points of the spectrum. For the easy-plane case ($g_x = g_y > g_z$), a biased spectral intensity with a derivative lineshape is expected on the low-field end of the spectrum (absorption onset), with a dip at the high-field end (cessation of the absorption). The low-field signal may be further biased in a loose powder due to torquing and preferential alignment of individual microcrystals. Therefore, we associate the strong derivative signal centered just above 2.0 T (frequency of 120.0 GHz) with $g_x$ and $g_y$ = 4.20(5). A weak dip corresponding to the $g_z$ component of the spectrum is harder to pick-out, as it rides on top of a broad signal spanning the $g$ = 2.00 region that we ascribe to paramagnetic contaminants; the sharp signal exactly at $g$ = 2.00, marked with an asterisk in Fig. \ref{fig:EPR}, is also assigned to an impurity. Nevertheless, the sharp dip at the $g_z$ = 1.93(2) position persists to high temperatures and is consistent with signals observed each time the sample was re-measured. Error bars were determined from the linewidths of the observed signals (peak-to-peak linewidth in the case of the $g_x$/$g_y$ signal). Finally, the fact that the resonance positions do not shift upon varying the temperature indicates that magnetic interactions do not influence the measurements, thus confirming that the EPR is in fact probing the isolated Yb sites in the diluted sample. By contrast, measurements performed on concentrated samples (100\% Yb$^{3+}$ in place of Lu$^{3+}$) revealed broad EPR signals with strongly temperature dependent effective $g$-values, significantly shifted from the free-ion values due to the Yb-Yb exchange.

 Our EPR results confirm the $xy$ anisotropy of the $g$-tensor in \ybge{}, expected from powder studies~\cite{hallas2016xy}, but does not agree quantitatively with previous determinations of the $g$-values from INS~\cite{hallas2016xy}. We attribute this disagreement to an intrinsic ambiguity in the fitting of the INS CEF data in Ref.~[\onlinecite{hallas2016xy}]. A similar ambiguity is likely to affect the determination of CEF parameters for other Yb-based materials. See
  Appendix \ref{sec:fitting:cef} for details.

With the single-ion properties determined, the exchange interactions ($J_1, J_2, J_3, J_4$) can next be obtained using the high-field spectrum. In a field of 3 T applied along $[1\bar{1}0]$, \ybge{} is in a field-polarized paramagnetic state and spin-wave excitations can be observed via INS. Due to the coarseness of the energy resolution of our measurement compared to the bandwidth, the dispersion of the excitations could not be reliably fit in the conventional way (as energy versus momentum relations), as we show in Appendix \ref{sec:dispersion}. Therefore, to constrain the exchange interactions, we instead fit the intensity profile in the $[hhl]$ plane at fixed energy-transfer (constant $E$), taking into account the energy dependent energy resolution, as shown in Fig.~\ref{fig:INS}. The expected intensity was computed using the model Eq.~(\ref{eqn:hamiltonian}) via linear spin-wave theory, with the $g$-tensor values fixed to those determined by EPR, [$g_{\pm} = 4.20(5)$ and $g_{z} = 1.93(2)$]. Results were averaged in a small window out the scattering plane to account for the finite detector size. To further constrain the exchange parameters, we included in the fit the high-temperature part of the specific heat ($5\K < T < 8\K$) computed theoretically via a numerical linked-cluster expansion (NLC)~\cite{tang2013short,Applegate2012a,Hayre2012}.  Our best fit exchange interactions are given in Table~\ref{tab:params}, with Fig.~\ref{fig:INS} showing the good agreement between the calculations and the data using the best-fit parameters.

\begin{table}[]
\begin{tabular}{llllllll}
\toprule
\multicolumn{2}{c}{Local} & \multicolumn{2}{c}{Global} &
\multicolumn{2}{c}{Global (Alt.)} & \multicolumn{2}{c}{Dual (Alt.)} \\ \midrule
$J_{zz}$ &= $+0.128(95)$    & $J_1$ &= $-0.01(6)$ & $J$ &= $-0.01(6)$ & $\tilde{J}$  &= $+0.35(4)$ \\
$J_{\pm}$ &= $+0.138(6)$    & $J_2$ &= $-0.44(4)$ & $K$ &= $-0.43(7)$ & $\tilde{K}$  &= $-0.08(3)$ \\
$J_{\pm\pm}$ &= $+0.044(24)$    & $J_3$ &= $-0.37(6)$ & $\Gamma$ &= $-0.37(6)$ & $\tilde{\Gamma}$  &= $-0.02(7)$ \\
$J_{z\pm}$ &= $-0.188(18)$    & $J_4$ &= $-0.02(13)$ & $D$ &= $-0.03(2)$ & $\tilde{D}$  &= $-0.27(9)$ \\
\bottomrule
\end{tabular}
\caption{Best fit exchange parameters (in meV) for \ybge{}, determined from fitting INS and $C_p$, in several different (equivalent) presentations: local~\cite{savary2012order}, global~\cite{ross2011quantum} and alternate global and dual global forms~\cite{rau2018yb}. Uncertainties in the last few digits are shown in parentheses.}
\label{tab:params}
\end{table}

\subsection{Discussion}

The determined exchange parameters place \ybge{} (Table \ref{tab:params}) very close to the (classical) boundary between the SFM and $\Gamma_5$ phases. They indicate that \ybge{} lies within the $\Gamma_5$ phase classically, with leading quantum corrections selecting the $\psi_3$ state. This is consistent with the magnetic structure below $T_{\text{N}}$, which was previously reported to be either $\psi_2$ or $\psi_3$~\cite{dun2015antiferromagnetic}. While the classical phase boundaries are known to shift due to quantum fluctuations~\cite{PhysRevLett.115.267208,rau2019magnon}, the $\Gamma_5$ phase is expected \emph{grow} due the presence of soft modes~\cite{PhysRevLett.115.267208}. We therefore do not expect quantum corrections to affect our assignment of \ybge{} to the $\Gamma_5$ phase. However, our assignment of \ybge{} to $\psi_3$ is more tentative, given the uncertainties in our parameters (see Table~\ref{tab:params}) and their proximity to the boundary between the $\psi_2$ and $\psi_3$ phases. How the boundary between $\psi_2$ and $\psi_3$ changes as one goes beyond the classical approximation is less clear. One might expect that $\psi_3$ may be further stabilized at the expense of the $\psi_2$ due to additional soft modes that appear near the SFM phase boundary for the former~\cite{yan2017theory,canalsdm1,canalsdm2,chern2010pyrochlore}. This expectation, combined with $\psi_3$ occupying more phase space near our best fit parameters, leads us to conclude that \ybge{} is more likely to be in the $\psi_3$ state.

By locating \ybge{} on the phase diagram, we confirm that changing the non-magnetic cation from Ti to Ge, which presumably alters the superexchange interactions by modifying distances and bond angles~\cite{rau2018yb}, is enough to push the Yb pyrochlores just across the SFM-$\Gamma_5$ phase boundary.  Yet, the titanate and germanate are otherwise extremely similar.  The close relationship between these compounds is apparent even in powder samples: despite the disparate ordered ground states, a striking similarity is observed in the \emph{powder averaged} zero field excitation spectra of the Yb pyrochlores as probed by INS~\cite{hallas2016universal}, with each material exhibiting a continuum of excitations.
\begin{figure}[!tb]
\includegraphics[width = \columnwidth]{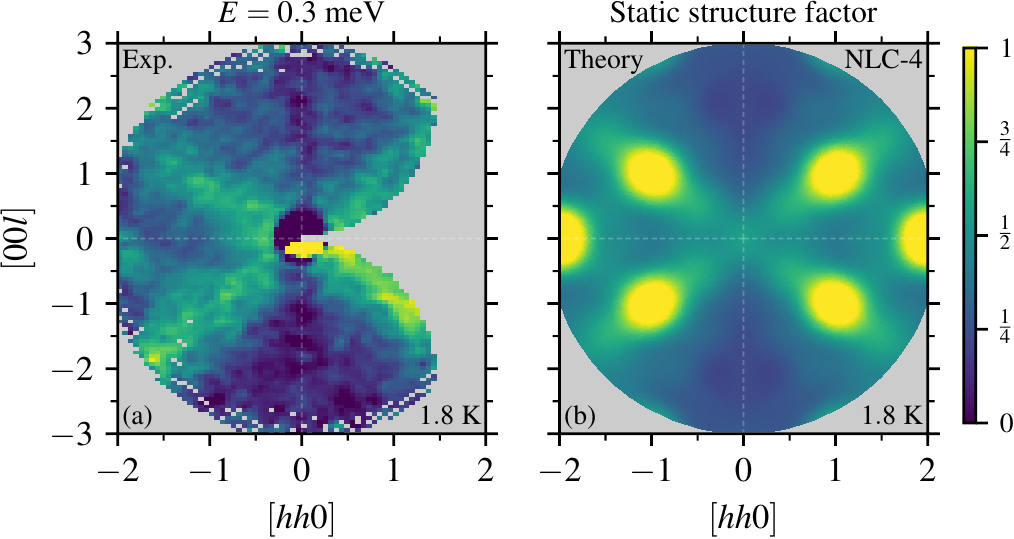}
\caption{(a) Constant energy transfer slice ($E = 0.3$ meV) for \ybge{} in zero field at $T=1.8\K$. ``Rods'' of scattering are visible along $\langle 111 \rangle$ directions as well as diffuse scattering at $(\bar{2}\bar{2}0)$. (b) Static (equal time) structure factor calculated using NLC at $T=1.8\K$ using our best fit parameters from the field-polarized spin waves (see Table~\ref{tab:params}), which also shows rods of scattering and intensity at $(220)$} 
\label{fig:zerofield}
\end{figure}

 One exciting potential explanation for the continuum is that the ordered phases in the\abo{Yb}{M} family are ``proximate'' to an exotic QSL brought on by the phase competition. 
Near such boundaries, classical degeneracies are enhanced, which can lead to the appearance of classical spin liquids~\cite{harris1997geometrical,bramwell2001spin, benton2016spin}. These are highly susceptible to quantum fluctuations~\cite{yan2017theory,PhysRevLett.115.267208} and can potentially help stabilize a QSL state~\cite{hermele2004pyrochlore,molavian2007,onoda2009,gingras2014quantum,liu2019competing}.
The effects of a nearby QSL in the Yb pyrochlores may explain unusual excitations such as the ones observed in powder samples~\cite{hallas2016universal}, as has been proposed for the  $\alpha$-RuCl$_3$ Kitaev material~\cite{banerjee2016proximate,banerjee2017neutron}. The nearby QSL phase may be accessible via the application of chemical or external pressure, or perhaps a combination of both, to the Yb pyrochlores.  In \ybti{} external hydrostatic pressure was found to further stabilize the FM state \cite{kermarrec_NatComm}. This suggests that external pressure on \ybge{} could move the compound in the same general direction, i.e. towards the SFM phase, and thus towards the phase boundary.

Zero-field spin excitations of single crystal \ybge{} for temperatures below $T_\text{N}$ collected at CNCS show broad and nearly featureless scattering, similar to polycrystalline INS reported in Ref.~[\onlinecite{hallas2016universal}] (Appendix \ref{sec:zerofield}).  The zero-field data collected at MACS, with an attributed temperature $T = 1.8 \K$, measures the correlations below the Schottky-like hump in the specific heat, a feature that coincides with the onset of significant structured paramagnetic scattering in \ybti{}~\cite{ross_rods,thompson_rods} and other Yb pyrochlores~\cite{hallas2018experimental} (as well as some reports of other quantum coherent phenomenon~\cite{tokiwa2016possible,pan2016measure}). We find that quasi-elastic paramagnetic scattering in \ybge{} at $0 \T$ and $1.8 \K$ qualitatively matches that of \ybti{} in the same regime; ``rods'' of scattering are observed along the $\langle 111 \rangle$ directions~\cite{ross_rods,thompson_rods}, with a broad peak near $(220)$ [Fig.~\subref{fig:zerofield}{(a)}]. A similar pattern is reproduced in the theoretical static structure factor computed via NLC (Fig.~\subref{fig:zerofield}{(b)} using our best fit parameters).  A detailed comparison of the zero-field excitations in \ybge{} and \ybti{} single crystals in their ordered states is worthy of future study.

\section{Conclusions}
In summary, we have presented single crystal neutron scattering data from \ybge{}, the sister compound of the well-studied pyrochlore \ybti{}. We have determined accurate values of the $g$-tensor of \ybge, measured directly by EPR spectroscopy of 1\% Yb-doped \luge{}. Fits to field-polarized INS and thermodynamic data, allow the determination of the four symmetry-allowed nearest neighbor exchanges, placing \ybge{} exquisitely close to the classical phase boundary between the SFM and $\Gamma_5$ phase, just inside the $\Gamma_5$ phase, with the leading quantum effects predicting a $\psi_3$ ground state. The zero field paramagnetic scattering in \ybge{} shows the same qualitative features as \ybti{}. Our work demonstrates the striking similarity between these two unconventional pyrochlores, and definitively locates \ybge{} on the phase diagram that has been so successful in describing a variety rare-earth pyrochlores~\cite{rau2019frustrated,yan2017theory,WongPRB2013}. Having established the proximity of \ybge{} to the SFM/$\Gamma_5$ boundary, and perhaps the $\psi_2$/$\psi_3$ one [see Fig.~\subref{fig:phasediag}{(a)}], one may now begin investigating how this affects the zero-field collective excitations of this compound~\cite{hallas2016universal}. Moreover, our work opens the door to tuning these Yb pyrochlores directly to the phase boundary, either by using external pressure~\cite{kermarrec_NatComm} or chemical pressure (e.g. YbTi$_{2-x}$Ge$_x$O$_7$). Finally, we have shown that relatively small single crystal samples obtained by hydrothermal synthesis can be used for detailed INS measurements, paving the way for other such measurements on crystals that can be grown using similar methods.




\begin{acknowledgements}
This research was partially supported by the CIFAR Quantum Materials program. KAR and CLS acknowledge support from the Department of Energy award DE-SC0020071 during the preparation of this manuscript.  Access to MACS was provided by the Center for High Resolution Neutron Scattering, a partnership between the National Institute of Standards and Technology and the National Science Foundation under Agreement No. DMR-1508249. This work was in part supported by Deutsche Forschungsgemeinschaft (DFG) under grant SFB 1143 and through the W\"urzburg-Dresden Cluster of Excellence on Complexity and Topology in Quantum Matter -- \textit{ct.qmat} (EXC 2147, project-id {390}{854}{90}). The work at
the University of Waterloo was supported by the Canada
Research Chair program (M.J.P.G., Tier 1). A portion of this work was performed at the National High Magnetic Field Laboratory, which is supported by the National Science Foundation Cooperative Agreement No. DMR-1644779 and the state of Florida.
\end{acknowledgements}
\appendix
\counterwithin{figure}{section}

\section{Magnetization}
\begin{figure*}
\includegraphics[width=0.8\textwidth]{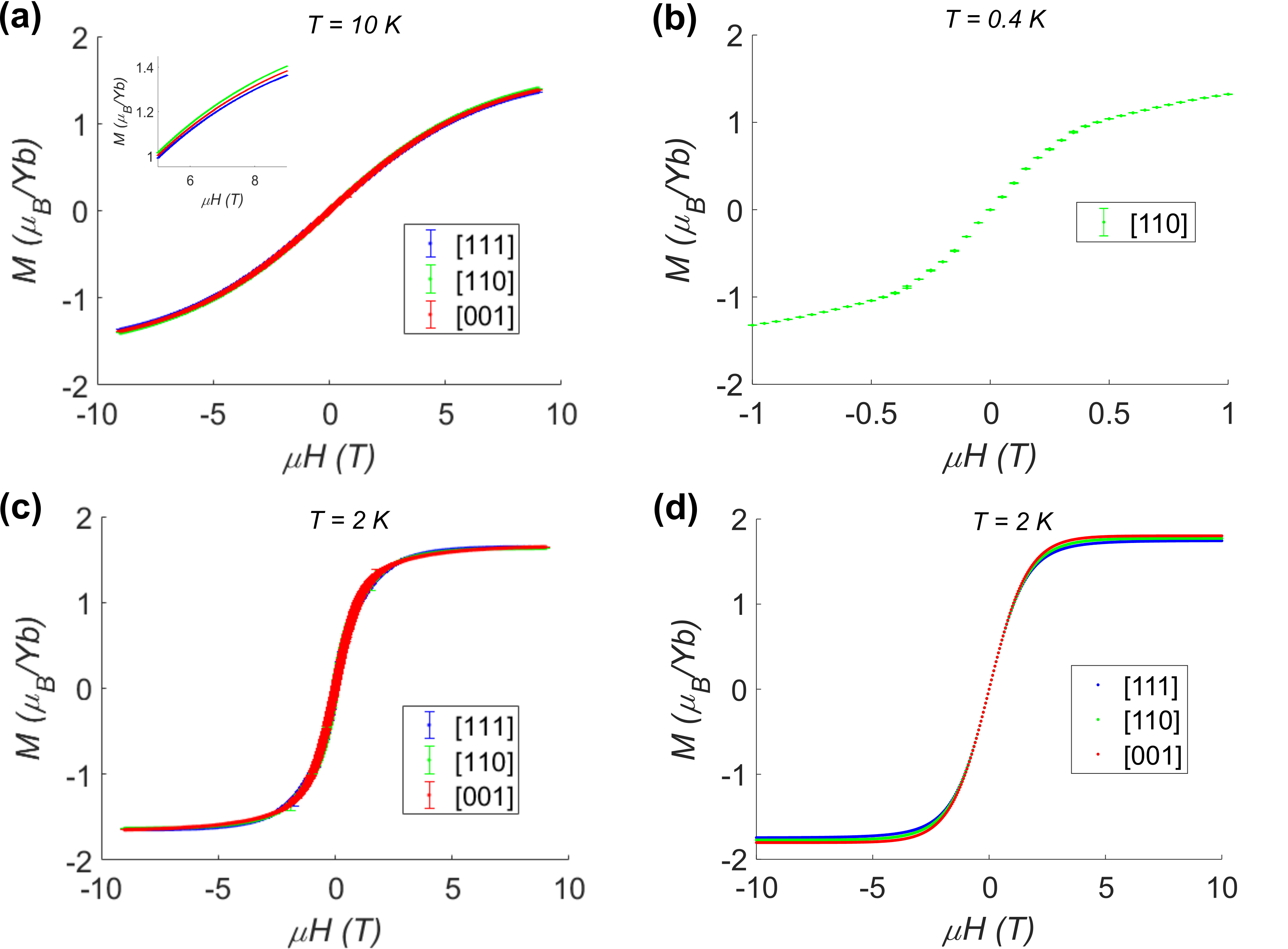}
\caption{Magnetization vs magnetic field for a \ybge\  single crystal. (\textit{a}) Data taken at $T=10$ K for three high symmetry directions of pyrochlore lattice. (\textit{b}) Data taken at $T=0.4$ K, with field along [110]. (\textit{c}) Data taken at $T=2$ K. (\textit{d}) Calculated single ion magnetization (using $g_z=1.93$, $g_{\pm}=4.20$) at $T=2$ K.} 
\label{fig:MvH}
\end{figure*}

Magnetization on a small single crystal ($m=0.68\ {\rm mg}$) of \ybge{} was performed using vibrating sample magnetometry (VSM) on a Quantum Design Dynacool PPMS.  Three separate measurements were performed such that the field was aligned with each of the high symmetry directions of the pyrochlore lattice ($[111]$, $[110]$, $[001]$). Correct orientation was checked prior to and after measurement to rule out sample movement during the measurements. Magnetization versus field curves show a nearly isotropic response at $T=2 \K$ and $10 \K$ [Fig.~\subref{fig:MvH}{(a,c)}].  The data are nearly in agreement with the expectations for the single-ion using the $g$-tensor values extracted from EPR [Fig.~\subref{fig:MvH}{(d)}]. We also note that the saturated moment at 2 K,  $\mu \sim 1.6\mu_{\rm B}$, agrees well with previous literature \cite{dun2015antiferromagnetic}.  The small deviations from the single ion model are likely attributable to the effect of exchange interactions, which are not negligible even relative to the maximum field strength.

Lower temperature (0.4 K) magnetization data with the field applied along [110] were collected using a Quantum Design MPMS SQUID magnetometer with ${}^3$He insert [Fig.~\subref{fig:MvH}{(b)}].  A kink in the magnetization data reveal a phase transition into the field-polarized paramagnetic state around 0.4 T. 

\section{Details of INS measurements}
\label{sec:INS}
\begin{figure}
\includegraphics[width=0.9\columnwidth]{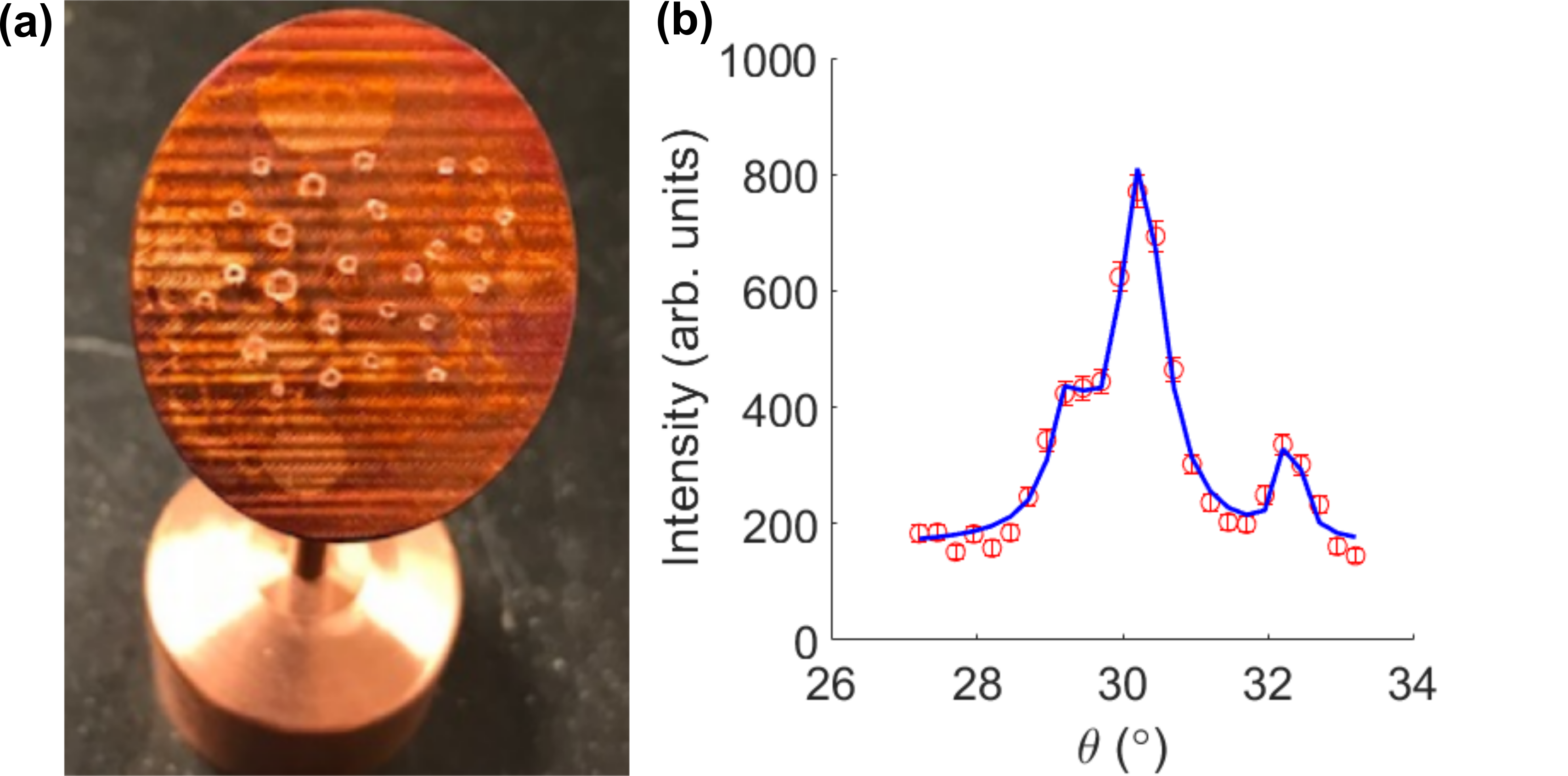}
\caption{(a): MACS neutron sample mount with 28 co-aligned single crystals of \ybge. (b):  Example rocking scan taken at MACS across a (111) nuclear peak at  $T\sim 6$ K.The peak is split into three peaks indicating a mosaic of $\leq 5^\circ$}.
\label{fig:rscan}
\end{figure}

Due to the difficulty in growing large (i.e., cm$^3$ sized) single crystals of the metastable pyrochlore phase of \ybge{}, we were restricted to small high-quality single crystals (1mm x 1mm x 1mm). To increase the sample volume for neutron scattering, we co-aligned 28 small single crystals in the $hhl$ scattering plane ([$1\bar{1}0$] direction vertical) for a total mass of 154 mg [Fig.~\subref{fig:rscan}{(a)}].   The crystals were fixed in place using a fluorinated glue (CYTOP 807-M). A rocking scan was taken over a  (111) nuclear peak, shown in Fig.~\subref{fig:rscan}{(b)}. We note a peak splitting consistent with a  mosaic of $\leq 5^\circ$ over all 28 crystals.

At MACS, INS data were taken throughout the $[hhl]$ plane at a constant energy transfer ($E = |E_{\rm f} - E_{\rm i}|$), using a fixed final energy of $E_{\rm f}=3.7$~meV and varying $E_{\rm i}$. The monochromator was used in doubly-focused mode with no radial collimators or filters in the incident beam, and cooled BeO filters were used in the scattered beam before the detectors. This configuration produces an energy resolution of 0.17 meV at the elastic line~\cite{MACSsite}. At each $E$ (which increased in 0.1 meV steps from 0 to 1.5 meV), the sample was rotated through 180$^{\circ}$ in 2$^{\circ}$ increments, counting for $1.66\times 10^5$ monitor units (approximately 10 s) at each increment.

As mentioned in the main text, data taken in zero field at the base temperature of the dilution refrigerator (mixing chamber temperature reading 260 mK) during the INS measurement at MACS is indistinguishable from data taken at 1.8 K.  One possibility is that the inelastic spectrum is basically insensitive to temperature below 1.8 K, which would be largely consistent with a previous powder study~\cite{hallas2016universal}.  However, in our experiment the elastic scattering also does not show any dependence on temperature below $T= 1.8$ K, even though it is clear that AFM Bragg peaks should develop below $T_{\rm N}$ (as has indeed been observed in the powder samples\cite{dun2015antiferromagnetic}). We can thus only conclude that the sample did not cool below $T_{\rm N}$, potentially due to the large mass of the sample holder
(necessary to hold the 28 co-aligned crystals), or weak thermal coupling between the small crystals and the sample holder. We thus assign a sample temperature of 1.8 K for our field-polarized INS data.

\begin{figure}
\includegraphics[width=0.9\columnwidth]{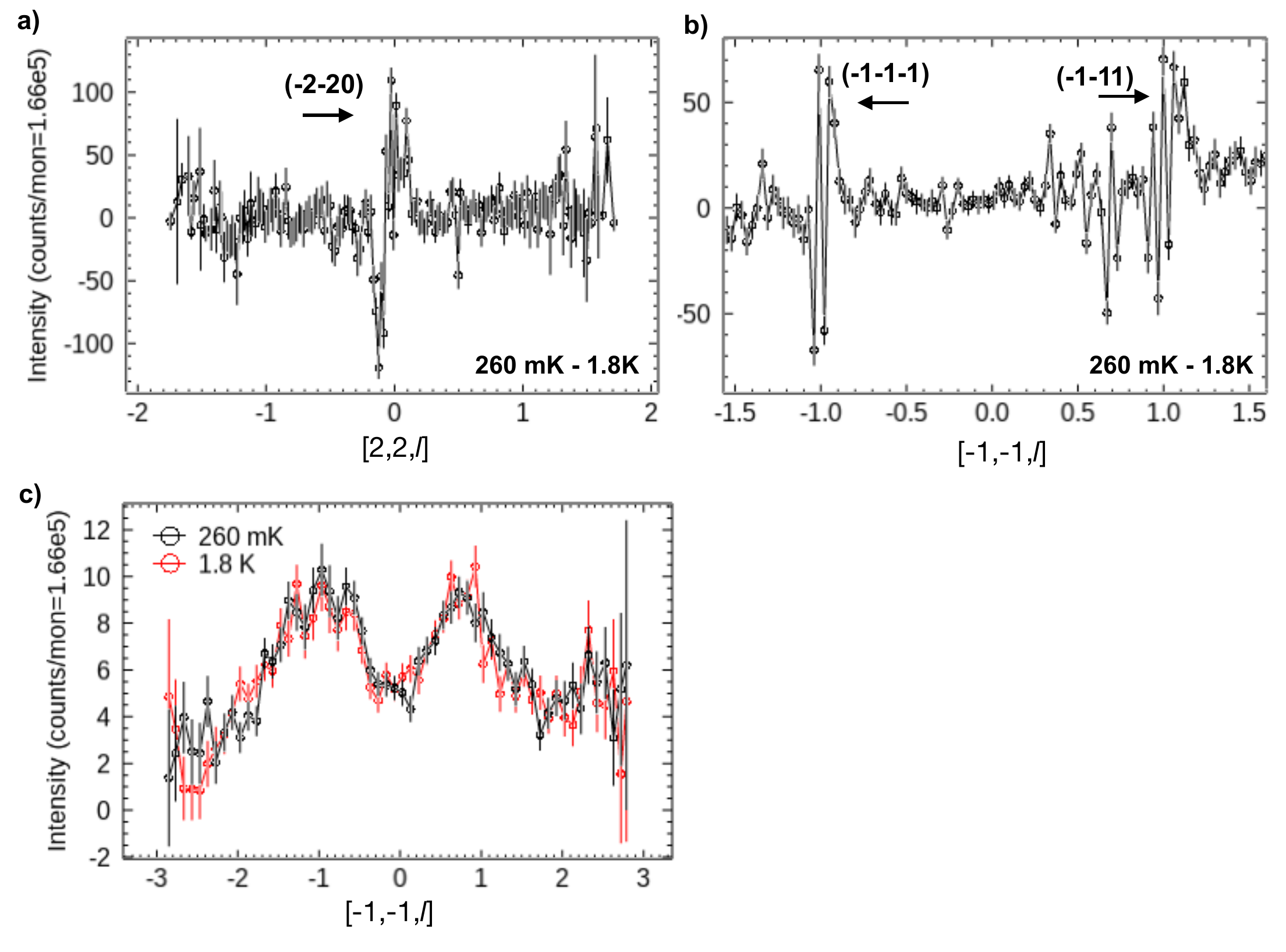}
\caption{Nominal temperature dependence of neutron scattering data taken on MACS.  (a) and (b) show a lack of additional intensity on expected AFM Bragg peak positions.  A shift of the peaks is observed instead (resulting in a net-zero profile with some regions of negative and positive differences). (c) overlay of the intensity along $[-1,-1,l]$ of the low energy excitations, $E = 0.3 \meV$, at both temperatures (3 T data used as background subtraction).}
\label{fig:temps}
\end{figure}

Figure \ref{fig:temps} shows the comparison between the $T = 260 \mK$ (nominal) and $T = 1.8\K$ data.

\subsection{Dispersion of 3 T data}
\label{sec:dispersion}
\begin{figure}
\includegraphics[width=0.7\columnwidth]{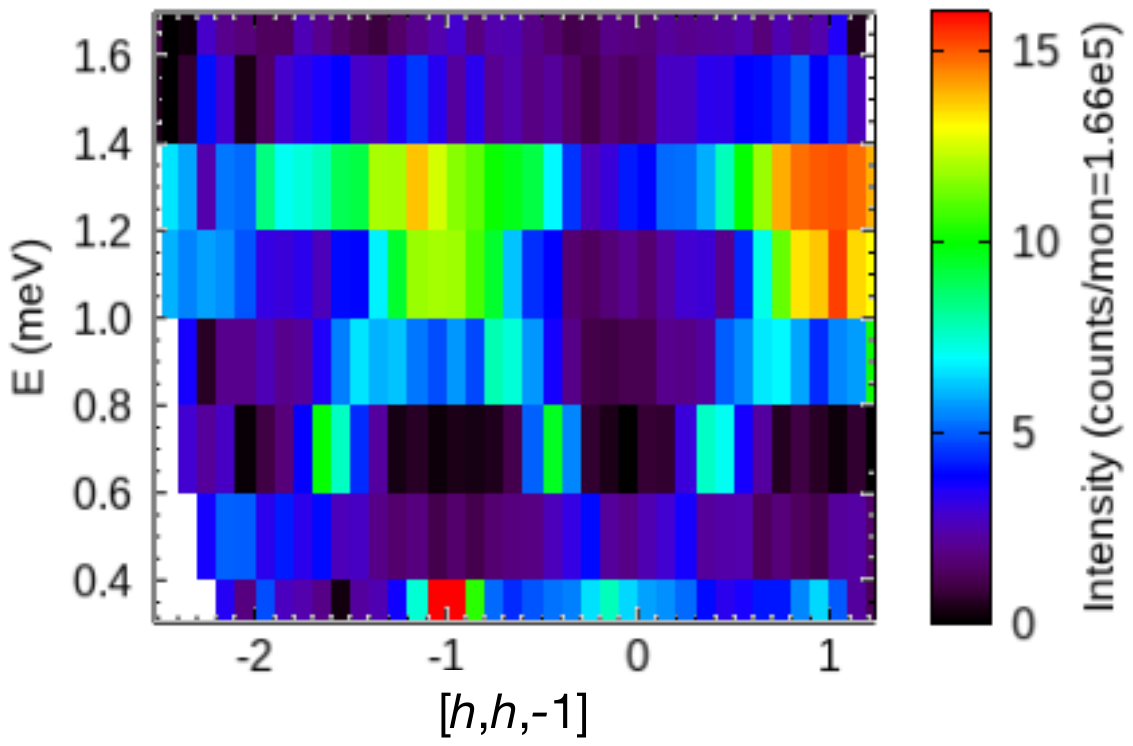}
\caption{\ybge{} INS data presented as a typical spin wave dispersion plot along $[h,h,$-1$]$. The dispersion is constructed from the combination of several constant energy slices through the $[hhl]$ plane.}
\label{fig:dispersion}
\end{figure}

Figure \ref{fig:dispersion} shows INS data presented as a typical spin wave dispersion plot, illustrating that the energy resolution is insufficient to uniquely resolve each spin-wave branch and thus it is not feasible to fit the dispersions themselves.  Instead, we fit the intensity at several energies over the whole $[hhl]$ plane.

\subsection{Dispersion of zero-field data}
\label{sec:zerofield}
\begin{figure*}
    \centering
    \includegraphics[width=\textwidth]{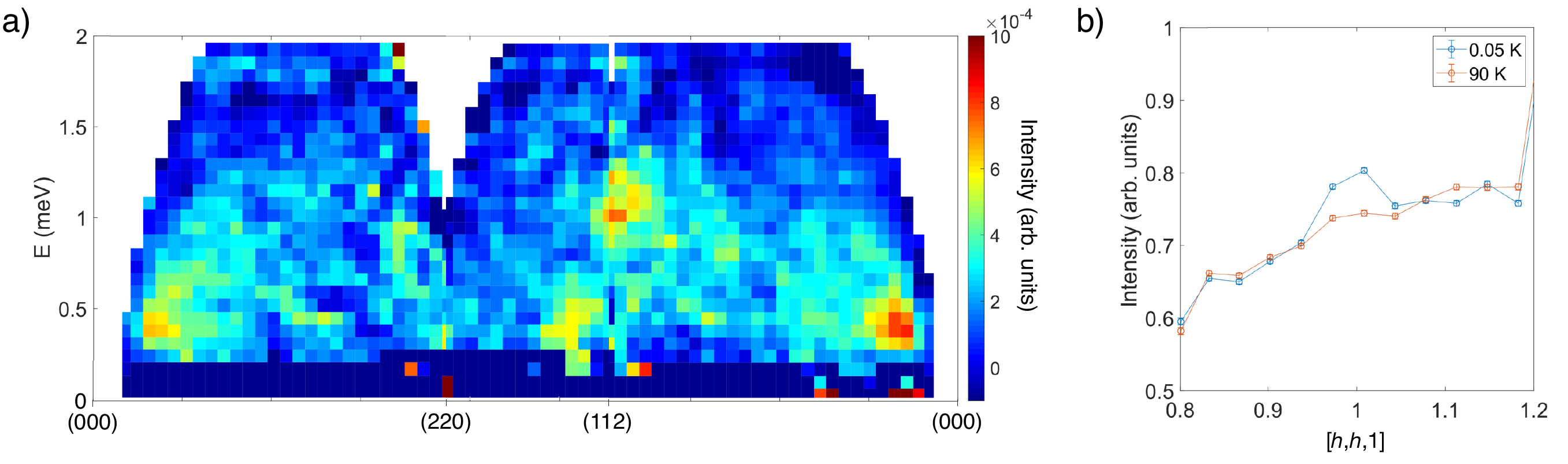}
    \caption{(a) Zero-field INS spectrum of \ybge\ at 60~mK, within the ordered state. All reciprocal space directions were integrated $\pm 0.1$ r.l.u. in their respective perpendicular components (for example, the cut along $[hh0]$ is integrated $\pm 0.1$ r.l.u. in $[00$l$])$. The excitations are broad and featureless, despite the ordered ground state, similar to powder INS data reported in Ref.~[\onlinecite{hallas2016universal}].(b) comparison of (111) Bragg peak at 60~mK and 90~K showing added intensity from magnetic ordering. Plots were integrated from $\pm 0.1$~meV and $\pm 0.1$ r.l.u. in $[00l]$.}
    \label{fig:CNCS}
\end{figure*}

A measurement of the zero-field, low temperature (60 mK) dynamic structure factor was performed on the cold neutron chopper spectrometer (CNCS) at the Spallation Neutron Source (SNS) at Oak Ridge National Laboratory (ORNL), using an 8T magnet (8T field-polarized data was used as background subtraction for zero-field data) with dilution refrigerator insert. The same crystals and crystal mount were used in this experiment as in the MACS experiment in the main text, oriented with the [1$\bar{1}$0] along the vertical field direction to access the horizontal $[hhl]$ scattering plane. Data was taken throughout the $[hhl]$ plane with an incident neutron energy of 
$E_{\rm i}$= 2.5~meV, operated in high-flux mode which gave an energy resolution of 0.07 meV at the elastic line~\cite{CNCSsite,CNCSressite}. The sample rotation method was used, where the sample was rotated 180$^\circ$ around the vertical in 2$^\circ$ steps. Data was collected at 0~T and 8~T (used for background subtraction) with data taken above the ordering transition (90~K) and below the ordering transition (60~mK reported at the mixing chamber). The clear temperature dependence of the (111) peak indicates that the sample cooled below the transition temperature (Fig. \subref{fig:CNCS}{(b)}).

Below the ordering transition, the excitation spectrum remains broad and featureless across reciprocal space, shown in Fig.~\subref{fig:CNCS}{(a)}. The bandwidth of the excitations is approximately 1.25 meV.  This is similar to polycrystalline INS data taken by Hallas {\it et  al.}~\cite{hallas2016universal}, who also reported broad diffuse dispersion below the ordering transition.

\section{Details of fitting}
\label{sec:fitting}
\subsection{Ambiguities in fitting crystal field parameters}
\label{sec:fitting:cef}
In this section, we address in more detail why it was necessary to obtain additional information above and beyond the inelastic neutron scattering data, specifically EPR on a diluted sample (as described in Sec.~\ref{sec:results:a}), to determine the $g$-factors.

A commonly used and reliable approach to finding the $g$-factors in a rare-earth magnet, such as \ybge{}, is through determination of the parameters that describe the crystal field potential via fitting to experimental data. Typically, in INS measurements, the data used would be the transition energies and intensities between different crystal field multiplets. However, two issues present themselves in \ybge{}. First, the energy scale of the crystal field is very large, with the gap between the ground and first excited doublet being $\sim 80 \meV$~\cite{hallas2016xy}. At accessible experimental temperatures, one can thus only probe transitions from the ground doublet (which is essentially fully populated) to the excited doublets (which are unpopulated).
Second, for the $D_{3\rm d}$ environment of Yb$^{3+}$ in \ybge{}, the $J=7/2$ manifold is split into only four Kramers doublets. Given the overall absolute intensity scale is (typically) difficult to determine, this leaves only \emph{five} pieces of information: three transition energies (ground to excited levels for the reason stated above) and two relative transition intensities. This is less than the \emph{six} parameters needed to describe the crystal field; usually denoted $B_{20}$, $B_{40}$, $B_{60}$, $B_{43}$, $B_{63}$ and $B_{66}$ (see, e.g. Ref.~[\onlinecite{bertin2012crystal}] for details). Such fitting, for example as carried out in Ref.~[\onlinecite{gaudet2015cef}] for \yto{} and in Ref.~[\onlinecite{hallas2016xy}] for \ybge{}, is thus underconstrained and generically cannot yield a unique best fit. We note that in some ytterbium magnets where the crystal field energy scale is smaller,  neutron data at several different temperatures may be used to resolve this issue (see, for example, Ref.~[\onlinecite{sala2019crystal}]).

\begin{figure}
    \centering
    \includegraphics[width=0.9\columnwidth]{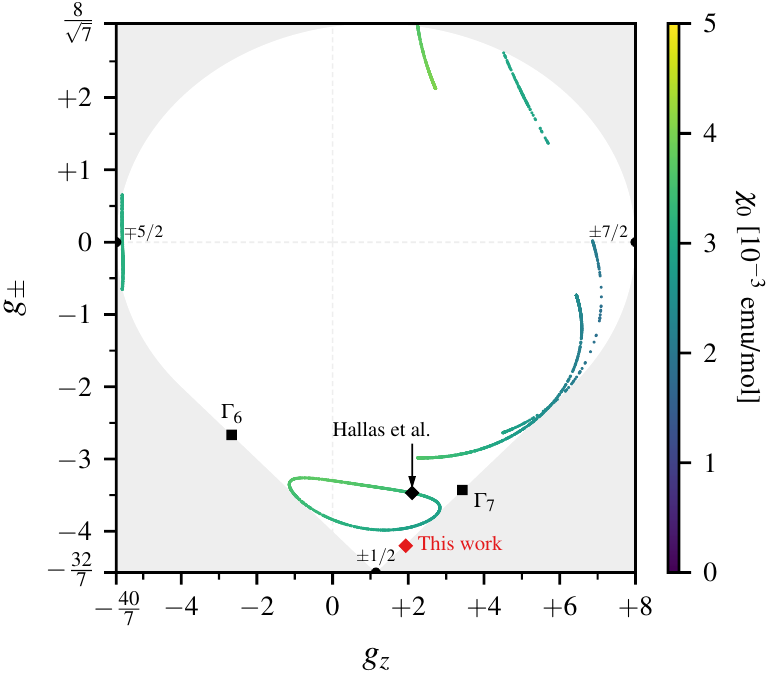}
    \caption{Illustration of the non-uniqueness of the $g$-factors obtained from the fit reported in Ref.~[\onlinecite{hallas2016xy}]. Each set of $g$-factors represents a set of six crystal field parameters ($B_{kq}$) with transition energies and relative intensities within $1\%$ of the result computed using the best fit parameters of Ref.~[\onlinecite{hallas2016xy}]. The color of each point show the variation of the Van Vleck contribution ($\chi_0$) to the susceptibility. Several important limits are indicated: $\Gamma_6$ doublet (octahedron cage), $\Gamma_7$ doublet (cube cage), as well as pure $\ket{\pm 1/2}$, $\ket{\pm 5/2}$ and $\ket{\pm 7/2}$ doublets. The $g$-factors in the shaded region are not physical for a pure $J=7/2$ manifold in a $D_{3\rm d}$ crystal field~\cite{rau2018yb}}.
    \label{fig:cef1}
\end{figure}

To make this point explicit, we have performed a re-analysis of the fitting results of Ref.~[\onlinecite{hallas2016xy}] to highlight the non-uniqueness of the fit. Due to some ambiguities due to a phonon subtraction near the crystal field levels, we do \emph{not} attempt to directly refit their intensity as a function of energy. Instead, we determine sets of crystal field parameters that can reproduce the best fit transition energies (ground to excited) and relative intensities (which can be calculated using the best fit CEF parameters of Ref.~[\onlinecite{hallas2016xy}]), to within $1\%$ accuracy. The result of this fitting is shown in Fig.~\ref{fig:cef1}, where one sees that a large number of crystal field parameters can produce nearly identical transitions and relative intensities as their best fit, but with wildly different $g$-factors. Indeed, the manifold of fits shown in Fig.~\ref{fig:cef1} is (piece-wise) one-dimensional, as one would expect when trying to fit six parameters with only five pieces of data. One thus cannot use inelastic neutron scattering data alone to determine the $g$-factors in \ybge{}. 

We stress here that the set of $g$-factors in Fig.~\ref{fig:cef1} does not exhaust all potential values of $g_z$ and $g_{\pm}$ relevant for \ybge{}, since the phonon subtraction leaves a reasonable amount of uncertainty in properly assigning some of the transition energies and relative intensities.  Indeed, our final $g$-factors, determined in the diluted sample via EPR (see Sec.~\ref{sec:results:a}), do not belong to the manifold of fitted $g$ parameters shown in Fig.~\ref{fig:cef1}.

We also explored a joint fit between the high temperature susceptibility and the crystal field data (three transitions and two relative intensities), as an alternative to the EPR from the diluted samples. These fits were inconclusive due to both the phonon subtraction issue discussed above, and some variability depending on the temperature range used in fitting the high-temperature susceptibility. We do note that, while not determinative, these results are somewhat consistent with the $g$-factors determined by EPR, i.e. $(g_z,g_{\pm}) = (1.93,4.20)$. However, the Curie constant that was obtained by fitting the susceptibility was somewhat insensitive to the aforementioned confounding factors and is consistent with the EPR value.

\subsection{Fitting of the exchange parameters}
\label{sec:fitting:exchange}
In this section, we describe our fitting methodology to determine the four exchange constants $J_{zz}$, $J_{\pm}$, $J_{\pm\pm}$ and $J_{z\pm}$ (or, equivalently, in the global or dual bases). Throughout, we fix the $g$-factors to the ones found from EPR, that is  $(g_z,g_{\pm}) = (1.93,4.20)$ (for details, see Sec.~\ref{sec:results:a}). We consider data from two independent controlled ``perturbative'' regimes (high magnetic field or high temperature) to determine these four exchange constants. 

\begin{figure}
    \centering
    \includegraphics[width=0.8\columnwidth]{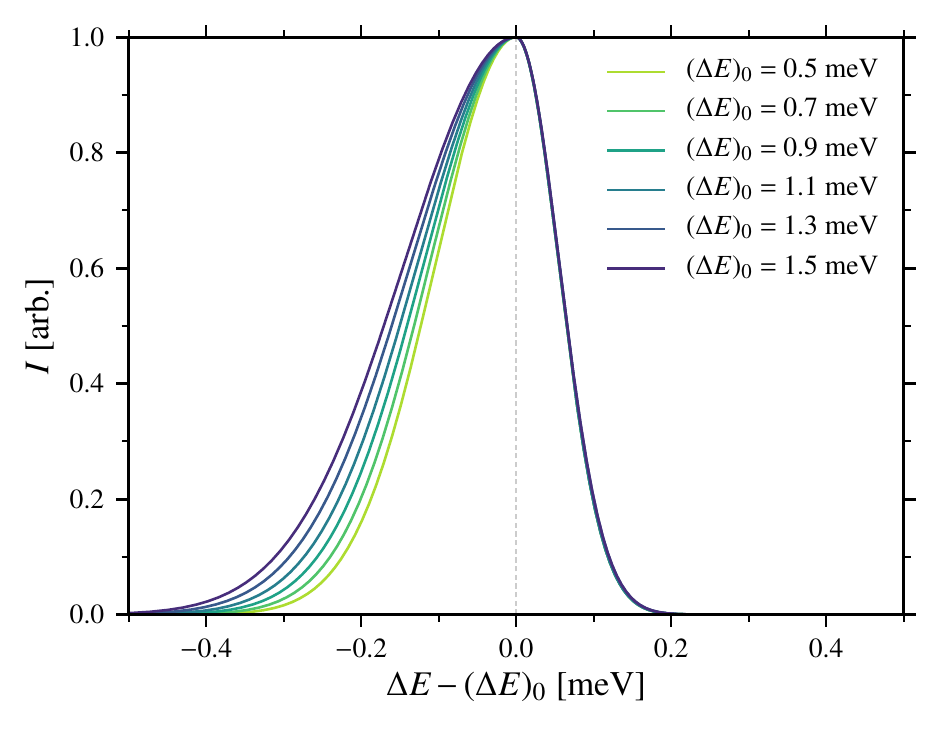}
    \caption{Energy resolution functions at energy transfer $\Delta E$ for specified energy transfers $(\Delta E)_0$ used in the theoretical calculations to emulate the behavior of the experimental setup. Practically these functions can be modelled by a two-sided Gaussian, with energy dependent width for $\Delta E < (\Delta E)_0$.}
    \label{fig:resolution}
\end{figure}

First, is the inelastic response in the high-field partially polarized phase obtained by applying a $B=3\T$ magnetic field along $[1\bar{1}0]$ at $T=1.8\K$. Theoretically, the inelastic response can be tractably calculated using standard linear spin-wave theory, as has been used in previous determinations of exchange constants in \yto{}~\cite{ross2011quantum,robert2015spin,thompson2017quasiparticle} and in \eto{}~\cite{savary2012order}. Due to experimental limitations (see Sec.~\ref{sec:dispersion}), instead of fitting the spin-wave \emph{spectrum}, we consider the inelastic intensity as a function of wave-vector (in the $[hhl]$ plane) within several fixed energy windows. Specifically, we consider the four energies $E = 0.5\meV, 0.7\meV, 0.9 \meV$ and $1.1\meV$ each averaged over an energy window with a resolution function that depends on the particular energy slice used (see Fig.~\ref{fig:resolution} for the precise form). To include the extent of the detectors out of the scattering plane, we also averaged over a window of $[-0.28,+0.28]$ r.l.u. in the $[1\bar{1}0]$ direction (the effect of finite resolution in the scattering plane is negligible for our purposes). The magnetic form factor of Yb$^{3+}$ was included in the fitting~\cite{inttablesc}, though the temperature is not, as the thermal population factors are unimportant even in the lowest energy window considered. Further, given we fit within (somewhat) narrow energy windows, any thermal factors primarily affect the overall intensity scale, not the variation with wave-vector.

Second, we make use of the specific heat data at zero field, but at high temperatures. This high temperature regime can be readily accessed using series expansion techniques. For this purpose, we employ a numerical linked-cluster expansion~\cite{nlc1,nlc2,tang2013short} to third order (NLC-3) in the number tetrahedra~\cite{Hayre2012,Applegate2012a,PhysRevLett.115.267208}. This order in the expansion is sufficient for good convergence in the temperature range considered for typical exchange constants, yet sufficiently fast computationally to still be amenable to automated fitting. More specifically, we consider five temperatures in the range $5\K \leq T \leq 8\K$ (to minimize any phonon effects) with the specific heat of the non-magnetic analog \abo{Lu}{Ge} subtracted~\cite{dun2015antiferromagnetic}.
\begin{figure*}
	\centering
	\includegraphics[width=\textwidth]{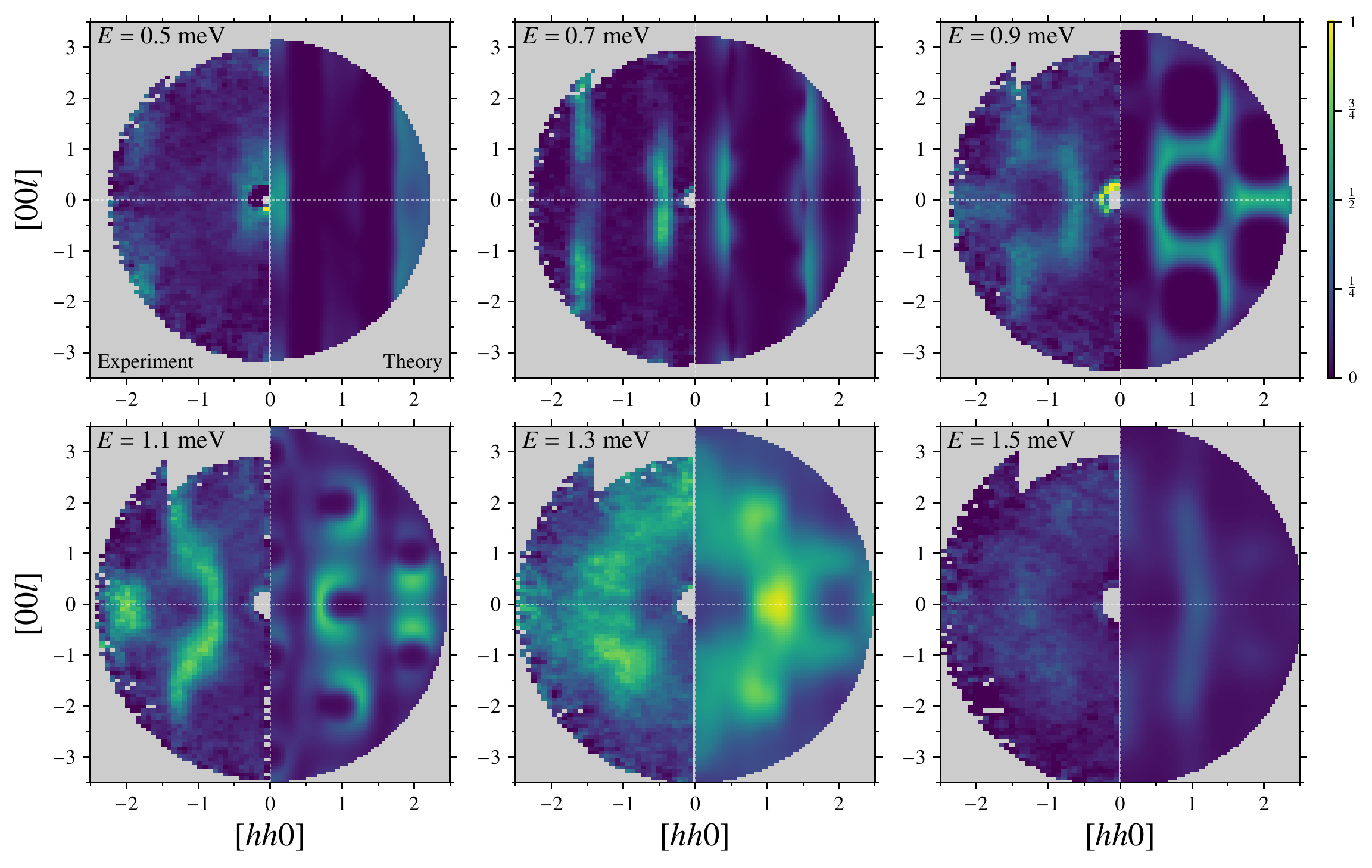}
	\caption{Comparison of constant-energy slices (centered at energy $E$ with an energy dependent energy resolution function, see Fig.~\ref{fig:resolution}) of the 3 T field polarized spin-waves between \ybge{} at $1.8\K$ (left) and linear spin wave theory using the best fit exchange parameters within linear spin-wave theory (right). Overall intensity scale is consistent between panels, but arbitrary. For the $1.3\meV$ and $1.5\meV$ cuts we assume the energy resolution function is the same as the $1.1\meV$ case. }
	\label{fig:high-energy}
\end{figure*}
For each comparison to experimental data (high-field inelastic response or high-temperature specific heat), we evaluate a $\chi^2$ value using estimates of the experimental errors, then sum these to obtain a total $\chi^2$ value. This total $\chi^2$ is then minimized to find the best fit (via standard Nelder-Mead simplex method implementation). To ensure that we find the global minimum, we repeat the fitting procedure $O(10^2)$ times from random initial points, typically taking each of the four exchanges $(J_{zz}, J_{\pm}, J_{\pm\pm}, J_{z\pm})$ to be chosen  independently and uniformly in the range $-0.2\meV$ to $+0.2\meV$. The best fit found is
\begin{subequations}
\begin{align}
    J_{zz} &= +0.128(95) \meV, \\
    J_{\pm} &= +0.138(6)\meV, \\
    J_{\pm\pm} &= +0.044(24) \meV, \\
    J_{z\pm} &= -0.188(18) \meV, 
\end{align}
\end{subequations}
as given in Table I of the main text. Statistical uncertainties are estimated using standard techniques by computing the curvature of $\chi^2$ about the best fit minimum. The uncertainties are most significant for the parameter $J_{zz}$. This is consistent with previous fits of experimental data in \yto{} have also found $J_{zz}$ to be constrained more loosely than the other fitted exchange parameters~\cite{thompson2017quasiparticle}. The curvature of $\chi^2$ provide the full (Gaussian) covariance matrix which was then transformed to obtain the corresponding error estimates for the global, alternate global and alternate dual exchanges presented in Table I of the main text.

We note that two high energy slices of the experimental data were not included in the fitting -- specifically those at $1.3\meV$ and $1.5\meV$. These were excluded for three reasons, first that they were mostly featureless and thus did not provide much additional information. Second that the intensity showed evidence of some spurious experimental effects, such as not following the known lattice symmetries (potentially due to differences in absorption for different rotation angles). Third, due to the greater importance of anharmonic magnon interactions (renormalization and spontaneous decay) at high energies, the applicability of linear spin-wave theory becomes questionable. The effects of these interactions is likely even an issue at $1.1 \meV$, as the one-magnon and two-magnon states overlap at this energy, and might account for some of the quantitative disagreements between the theory and experiment at this energy.

However, for completeness we include a comparison between the theory and experiment at all energies, as shown in Fig.~\ref{fig:high-energy}. It can be seen that while there are some differences, there is broad qualitative agreement for both the $1.3 \meV$ and $1.5\meV$ energy slices.

Finally, we note that we also attempted determining the $g$-factors without the EPR data, using only the high-field inelastic neutron scattering data and the Curie constant obtained from the magnetic susceptibility. Including the Curie constant fixes $\bar{g}^2 \equiv g_z^2+2g_{\pm}^2$, giving the pair of $g$-factors in terms of a single angle, $\theta$, as $g_z = \bar{g}\cos{\theta}$ and $g_{\pm} = \bar{g}\sin{\theta}/\sqrt{2}$. The fitting procedure described above was then carried out on the high-field inelastic neutron scattering data alone, on a grid to determine the value of $\theta$ having the lowest $\chi^2$. This was inconclusive, given the issues described above with temperature range dependence in determining the Curie constant and a large number of nearly equally good local minima. However, we do note that the $g$-factor values obtained via the EPR measurements is present among these local minima determined using susceptibility data.

%
%

%


\begin{thebibliography}{83}%
	\makeatletter
	\providecommand \@ifxundefined [1]{%
		\@ifx{#1\undefined}
	}%
	\providecommand \@ifnum [1]{%
		\ifnum #1\expandafter \@firstoftwo
		\else \expandafter \@secondoftwo
		\fi
	}%
	\providecommand \@ifx [1]{%
		\ifx #1\expandafter \@firstoftwo
		\else \expandafter \@secondoftwo
		\fi
	}%
	\providecommand \natexlab [1]{#1}%
	\providecommand \enquote  [1]{``#1''}%
	\providecommand \bibnamefont  [1]{#1}%
	\providecommand \bibfnamefont [1]{#1}%
	\providecommand \citenamefont [1]{#1}%
	\providecommand \href@noop [0]{\@secondoftwo}%
	\providecommand \href [0]{\begingroup \@sanitize@url \@href}%
	\providecommand \@href[1]{\@@startlink{#1}\@@href}%
	\providecommand \@@href[1]{\endgroup#1\@@endlink}%
	\providecommand \@sanitize@url [0]{\catcode `\\12\catcode `\$12\catcode
		`\&12\catcode `\#12\catcode `\^12\catcode `\_12\catcode `\%12\relax}%
	\providecommand \@@startlink[1]{}%
	\providecommand \@@endlink[0]{}%
	\providecommand \url  [0]{\begingroup\@sanitize@url \@url }%
	\providecommand \@url [1]{\endgroup\@href {#1}{\urlprefix }}%
	\providecommand \urlprefix  [0]{URL }%
	\providecommand \Eprint [0]{\href }%
	\providecommand \doibase [0]{http://dx.doi.org/}%
	\providecommand \selectlanguage [0]{\@gobble}%
	\providecommand \bibinfo  [0]{\@secondoftwo}%
	\providecommand \bibfield  [0]{\@secondoftwo}%
	\providecommand \translation [1]{[#1]}%
	\providecommand \BibitemOpen [0]{}%
	\providecommand \bibitemStop [0]{}%
	\providecommand \bibitemNoStop [0]{.\EOS\space}%
	\providecommand \EOS [0]{\spacefactor3000\relax}%
	\providecommand \BibitemShut  [1]{\csname bibitem#1\endcsname}%
	\let\auto@bib@innerbib\@empty
	\bibitem [{\citenamefont {Grissonnanche}\ \emph {et~al.}(2014)\citenamefont
		{Grissonnanche}, \citenamefont {Cyr-Choini{\`e}re}, \citenamefont
		{Lalibert{\'e}}, \citenamefont {De~Cotret}, \citenamefont {Juneau-Fecteau},
		\citenamefont {Dufour-Beaus{\'e}jour}, \citenamefont {Delage}, \citenamefont
		{LeBoeuf}, \citenamefont {Chang}, \citenamefont {Ramshaw} \emph
		{et~al.}}]{grissonnanche2014direct}%
	\BibitemOpen
	\bibfield  {author} {\bibinfo {author} {\bibfnamefont {G.}~\bibnamefont
			{Grissonnanche}}, \bibinfo {author} {\bibfnamefont {O.}~\bibnamefont
			{Cyr-Choini{\`e}re}}, \bibinfo {author} {\bibfnamefont {F.}~\bibnamefont
			{Lalibert{\'e}}}, \bibinfo {author} {\bibfnamefont {S.~R.}\ \bibnamefont
			{De~Cotret}}, \bibinfo {author} {\bibfnamefont {A.}~\bibnamefont
			{Juneau-Fecteau}}, \bibinfo {author} {\bibfnamefont {S.}~\bibnamefont
			{Dufour-Beaus{\'e}jour}}, \bibinfo {author} {\bibfnamefont {M.-E.}\
			\bibnamefont {Delage}}, \bibinfo {author} {\bibfnamefont {D.}~\bibnamefont
			{LeBoeuf}}, \bibinfo {author} {\bibfnamefont {J.}~\bibnamefont {Chang}},
		\bibinfo {author} {\bibfnamefont {B.~J.}\ \bibnamefont {Ramshaw}},  \emph
		{et~al.},\ }\href@noop {} {\bibfield  {journal} {\bibinfo  {journal} {Nature
				Communications}\ }\textbf {\bibinfo {volume} {5}},\ \bibinfo {pages} {3280}
		(\bibinfo {year} {2014})}\BibitemShut {NoStop}%
	\bibitem [{\citenamefont {Dagotto}(2005)}]{dagotto2005complexity}%
	\BibitemOpen
	\bibfield  {author} {\bibinfo {author} {\bibfnamefont {E.}~\bibnamefont
			{Dagotto}},\ }\href@noop {} {\bibfield  {journal} {\bibinfo  {journal}
			{Science}\ }\textbf {\bibinfo {volume} {309}},\ \bibinfo {pages} {257}
		(\bibinfo {year} {2005})}\BibitemShut {NoStop}%
	\bibitem [{\citenamefont {Savary}\ and\ \citenamefont
		{Balents}(2016)}]{savary2016quantum}%
	\BibitemOpen
	\bibfield  {author} {\bibinfo {author} {\bibfnamefont {L.}~\bibnamefont
			{Savary}}\ and\ \bibinfo {author} {\bibfnamefont {L.}~\bibnamefont
			{Balents}},\ }\href@noop {} {\bibfield  {journal} {\bibinfo  {journal}
			{Reports on Progress in Physics}\ }\textbf {\bibinfo {volume} {80}},\
		\bibinfo {pages} {016502} (\bibinfo {year} {2016})}\BibitemShut {NoStop}%
	\bibitem [{\citenamefont {Lacroix}\ \emph {et~al.}(2011)\citenamefont
		{Lacroix}, \citenamefont {Mendels},\ and\ \citenamefont {Mila}}]{HFM_book}%
	\BibitemOpen
	\bibfield  {author} {\bibinfo {author} {\bibfnamefont {C.}~\bibnamefont
			{Lacroix}}, \bibinfo {author} {\bibfnamefont {P.}~\bibnamefont {Mendels}}, \
		and\ \bibinfo {author} {\bibfnamefont {F.}~\bibnamefont {Mila}},\ }\href@noop
	{} {\emph {\bibinfo {title} {Introduction to Frustrated Magnetism}}}\
	(\bibinfo  {publisher} {Springer-Verlag, Berlin, Heidelberg},\ \bibinfo
	{year} {2011})\BibitemShut {NoStop}%
	\bibitem [{\citenamefont {Chandra}\ and\ \citenamefont
		{Doucot}(1988)}]{chandra1988possible}%
	\BibitemOpen
	\bibfield  {author} {\bibinfo {author} {\bibfnamefont {P.}~\bibnamefont
			{Chandra}}\ and\ \bibinfo {author} {\bibfnamefont {B.}~\bibnamefont
			{Doucot}},\ }\href@noop {} {\bibfield  {journal} {\bibinfo  {journal}
			{Physical Review B}\ }\textbf {\bibinfo {volume} {38}},\ \bibinfo {pages}
		{9335} (\bibinfo {year} {1988})}\BibitemShut {NoStop}%
	\bibitem [{\citenamefont {Capriotti}\ and\ \citenamefont
		{Sorella}(2000)}]{capriotti2000spontaneous}%
	\BibitemOpen
	\bibfield  {author} {\bibinfo {author} {\bibfnamefont {L.}~\bibnamefont
			{Capriotti}}\ and\ \bibinfo {author} {\bibfnamefont {S.}~\bibnamefont
			{Sorella}},\ }\href@noop {} {\bibfield  {journal} {\bibinfo  {journal}
			{Physical Review Letters}\ }\textbf {\bibinfo {volume} {84}},\ \bibinfo
		{pages} {3173} (\bibinfo {year} {2000})}\BibitemShut {NoStop}%
	\bibitem [{\citenamefont {Cabra}\ \emph {et~al.}(2011)\citenamefont {Cabra},
		\citenamefont {Lamas},\ and\ \citenamefont {Rosales}}]{cabra2011quantum}%
	\BibitemOpen
	\bibfield  {author} {\bibinfo {author} {\bibfnamefont {D.~C.}\ \bibnamefont
			{Cabra}}, \bibinfo {author} {\bibfnamefont {C.~A.}\ \bibnamefont {Lamas}}, \
		and\ \bibinfo {author} {\bibfnamefont {H.~D.}\ \bibnamefont {Rosales}},\
	}\href@noop {} {\bibfield  {journal} {\bibinfo  {journal} {Physical Review
				B}\ }\textbf {\bibinfo {volume} {83}},\ \bibinfo {pages} {094506} (\bibinfo
		{year} {2011})}\BibitemShut {NoStop}%
	\bibitem [{\citenamefont {Reuther}\ \emph {et~al.}(2011)\citenamefont
		{Reuther}, \citenamefont {Abanin},\ and\ \citenamefont
		{Thomale}}]{reuther2011magnetic}%
	\BibitemOpen
	\bibfield  {author} {\bibinfo {author} {\bibfnamefont {J.}~\bibnamefont
			{Reuther}}, \bibinfo {author} {\bibfnamefont {D.~A.}\ \bibnamefont {Abanin}},
		\ and\ \bibinfo {author} {\bibfnamefont {R.}~\bibnamefont {Thomale}},\
	}\href@noop {} {\bibfield  {journal} {\bibinfo  {journal} {Physical Review
				B}\ }\textbf {\bibinfo {volume} {84}},\ \bibinfo {pages} {014417} (\bibinfo
		{year} {2011})}\BibitemShut {NoStop}%
	\bibitem [{\citenamefont {Gong}\ \emph {et~al.}(2015)\citenamefont {Gong},
		\citenamefont {Zhu},\ and\ \citenamefont {Sheng}}]{gong2015quantum}%
	\BibitemOpen
	\bibfield  {author} {\bibinfo {author} {\bibfnamefont {S.-S.}\ \bibnamefont
			{Gong}}, \bibinfo {author} {\bibfnamefont {W.}~\bibnamefont {Zhu}}, \ and\
		\bibinfo {author} {\bibfnamefont {D.~N.}\ \bibnamefont {Sheng}},\ }\href@noop
	{} {\bibfield  {journal} {\bibinfo  {journal} {Physical Review B}\ }\textbf
		{\bibinfo {volume} {92}},\ \bibinfo {pages} {195110} (\bibinfo {year}
		{2015})}\BibitemShut {NoStop}%
	\bibitem [{\citenamefont {Rau}\ and\ \citenamefont
		{Gingras}(2018)}]{rau2018yb}%
	\BibitemOpen
	\bibfield  {author} {\bibinfo {author} {\bibfnamefont {J.~G.}\ \bibnamefont
			{Rau}}\ and\ \bibinfo {author} {\bibfnamefont {M.~J.~P.}\ \bibnamefont
			{Gingras}},\ }\href@noop {} {\bibfield  {journal} {\bibinfo  {journal}
			{Physical Review B}\ }\textbf {\bibinfo {volume} {98}},\ \bibinfo {pages}
		{054408} (\bibinfo {year} {2018})}\BibitemShut {NoStop}%
	\bibitem [{\citenamefont {Hester}\ \emph {et~al.}(2019)\citenamefont {Hester},
	\citenamefont {Nair}, \citenamefont {Reeder}, \citenamefont {Yahne},
	\citenamefont {DeLazzer}, \citenamefont {Berges}, \citenamefont {Ziat},
	\citenamefont {Neilson}, \citenamefont {Aczel}, \citenamefont {Sala},\citenamefont {Quilliam},\citenamefont {Ross} }]{hester2019novel}%
\BibitemOpen
\bibfield  {author} {\bibinfo {author} {\bibfnamefont {G.}~\bibnamefont
		{Hester}}, \bibinfo {author} {\bibfnamefont {H.~S.}~\bibnamefont {Nair}},
	\bibinfo {author} {\bibfnamefont {T.~R.}~\bibnamefont {Reeder}}, \bibinfo
	{author} {\bibfnamefont {D.~R.}~\bibnamefont {Yahne}}, \bibinfo {author}
	{\bibfnamefont {T.~N.}~\bibnamefont {DeLazzer}}, \bibinfo {author}
	{\bibfnamefont {L.}~\bibnamefont {Berges}}, \bibinfo {author} {\bibfnamefont
		{D.}~\bibnamefont {Ziat}}, \bibinfo {author} {\bibfnamefont {J.~R.}~\bibnamefont
		{Neilson}}, \bibinfo {author} {\bibfnamefont {A.~A.}\ \bibnamefont {Aczel}},
	\bibinfo {author} {\bibfnamefont {G.}~\bibnamefont {Sala}}, \bibinfo {author} {\bibfnamefont {J.~A.}~\bibnamefont {Quilliam}}, \bibinfo {author} {\bibfnamefont {K.~A.}~\bibnamefont {Ross}},\
}\href@noop {} {\bibfield  {journal} {\bibinfo  {journal} {Physical review
			letters}\ }\textbf {\bibinfo {volume} {123}},\ \bibinfo {pages} {027201}
	(\bibinfo {year} {2019})}\BibitemShut {NoStop}%
	\bibitem [{\citenamefont {Bordelon}\ \emph {et~al.}(2019)\citenamefont
		{Bordelon}, \citenamefont {Kenney}, \citenamefont {Liu}, \citenamefont
		{Hogan}, \citenamefont {Posthuma}, \citenamefont {Kavand}, \citenamefont
		{Lyu}, \citenamefont {Sherwin}, \citenamefont {Butch}, \citenamefont {Brown}
		\emph {et~al.}}]{bordelon2019field}%
	\BibitemOpen
	\bibfield  {author} {\bibinfo {author} {\bibfnamefont {M.~M.}\ \bibnamefont
			{Bordelon}}, \bibinfo {author} {\bibfnamefont {E.}~\bibnamefont {Kenney}},
		\bibinfo {author} {\bibfnamefont {C.}~\bibnamefont {Liu}}, \bibinfo {author}
		{\bibfnamefont {T.}~\bibnamefont {Hogan}}, \bibinfo {author} {\bibfnamefont
			{L.}~\bibnamefont {Posthuma}}, \bibinfo {author} {\bibfnamefont
			{M.}~\bibnamefont {Kavand}}, \bibinfo {author} {\bibfnamefont
			{Y.}~\bibnamefont {Lyu}}, \bibinfo {author} {\bibfnamefont {M.}~\bibnamefont
			{Sherwin}}, \bibinfo {author} {\bibfnamefont {N.~P.}\ \bibnamefont {Butch}},
		\bibinfo {author} {\bibfnamefont {C.}~\bibnamefont {Brown}},  \emph
		{et~al.},\ }\href@noop {} {\bibfield  {journal} {\bibinfo  {journal} {Nature
				Physics}\ }\textbf {\bibinfo {volume} {15}},\ \bibinfo {pages} {1058}
		(\bibinfo {year} {2019})}\BibitemShut {NoStop}%
	\bibitem [{\citenamefont {Ranjith}\ \emph
	{et~al.}(2019{\natexlab{a}})\citenamefont {Ranjith}, \citenamefont
	{Dmytriieva}, \citenamefont {Khim}, \citenamefont {Sichelschmidt},
	\citenamefont {Luther}, \citenamefont {Ehlers}, \citenamefont {Yasuoka},
	\citenamefont {Wosnitza}, \citenamefont {Tsirlin}, \citenamefont {K{\"u}hne},  \citenamefont {Baenitz}
	\emph {et~al.}}]{ranjith2019field}%
\BibitemOpen
\bibfield  {author} {\bibinfo {author} {\bibfnamefont {K.~M.}~\bibnamefont
		{Ranjith}}, \bibinfo {author} {\bibfnamefont {D.}~\bibnamefont {Dmytriieva}},
	\bibinfo {author} {\bibfnamefont {S.}~\bibnamefont {Khim}}, \bibinfo {author}
	{\bibfnamefont {J.}~\bibnamefont {Sichelschmidt}}, \bibinfo {author}
	{\bibfnamefont {S.}~\bibnamefont {Luther}}, \bibinfo {author} {\bibfnamefont
		{D.}~\bibnamefont {Ehlers}}, \bibinfo {author} {\bibfnamefont
		{H.}~\bibnamefont {Yasuoka}}, \bibinfo {author} {\bibfnamefont
		{J.}~\bibnamefont {Wosnitza}}, \bibinfo {author} {\bibfnamefont {A.~A.}\
		\bibnamefont {Tsirlin}}, \bibinfo {author} {\bibfnamefont {H.}~\bibnamefont
		{K{\"u}hne}}, \bibinfo {author} {\bibfnamefont {M.}~\bibnamefont {Baenitz}},\ }\href@noop {} {\bibfield  {journal}
	{\bibinfo  {journal} {Physical Review B}\ }\textbf {\bibinfo {volume} {99}},\
	\bibinfo {pages} {180401(R)} (\bibinfo {year} {2019}{\natexlab{a}})}\BibitemShut
{NoStop}%
	\bibitem [{\citenamefont {Wu}\ \emph {et~al.}(2019)\citenamefont {Wu},
		\citenamefont {Nikitin}, \citenamefont {Wang}, \citenamefont {Zhu},
		\citenamefont {Batista}, \citenamefont {Tsvelik}, \citenamefont {Samarakoon},
		\citenamefont {Tennant}, \citenamefont {Brando}, \citenamefont {Vasylechko}
		\emph {et~al.}}]{wu2019tomonaga}%
	\BibitemOpen
	\bibfield  {author} {\bibinfo {author} {\bibfnamefont {L.}~\bibnamefont
			{Wu}}, \bibinfo {author} {\bibfnamefont {S.}~\bibnamefont {Nikitin}},
		\bibinfo {author} {\bibfnamefont {Z.}~\bibnamefont {Wang}}, \bibinfo {author}
		{\bibfnamefont {W.}~\bibnamefont {Zhu}}, \bibinfo {author} {\bibfnamefont
			{C.}~\bibnamefont {Batista}}, \bibinfo {author} {\bibfnamefont
			{A.}~\bibnamefont {Tsvelik}}, \bibinfo {author} {\bibfnamefont
			{A.}~\bibnamefont {Samarakoon}}, \bibinfo {author} {\bibfnamefont
			{D.}~\bibnamefont {Tennant}}, \bibinfo {author} {\bibfnamefont
			{M.}~\bibnamefont {Brando}}, \bibinfo {author} {\bibfnamefont
			{L.}~\bibnamefont {Vasylechko}},  \emph {et~al.},\ }\href@noop {} {\bibfield
		{journal} {\bibinfo  {journal} {Nature communications}\ }\textbf {\bibinfo
			{volume} {10}},\ \bibinfo {pages} {1} (\bibinfo {year} {2019})}\BibitemShut
	{NoStop}%
	\bibitem [{\citenamefont {Ranjith}\ \emph
	{et~al.}(2019{\natexlab{b}})\citenamefont {Ranjith}, \citenamefont {Luther},
	\citenamefont {Reimann}, \citenamefont {Schmidt}, \citenamefont {Schlender},
	\citenamefont {Sichelschmidt}, \citenamefont {Yasuoka}, \citenamefont
	{Strydom}, \citenamefont {Skourski}, \citenamefont {Wosnitza} \emph
	{et~al.}}]{ranjith2019anisotropic}%
\BibitemOpen
\bibfield  {author} {\bibinfo {author} {\bibfnamefont {K.~M.}~\bibnamefont
		{Ranjith}}, \bibinfo {author} {\bibfnamefont {S.}~\bibnamefont {Luther}},
	\bibinfo {author} {\bibfnamefont {T.}~\bibnamefont {Reimann}}, \bibinfo
	{author} {\bibfnamefont {B.}~\bibnamefont {Schmidt}}, \bibinfo {author}
	{\bibfnamefont {P.}~\bibnamefont {Schlender}}, \bibinfo {author}
	{\bibfnamefont {J.}~\bibnamefont {Sichelschmidt}}, \bibinfo {author}
	{\bibfnamefont {H.}~\bibnamefont {Yasuoka}}, \bibinfo {author} {\bibfnamefont
		{A.~M.}~\bibnamefont {Strydom}}, \bibinfo {author} {\bibfnamefont
		{Y.}~\bibnamefont {Skourski}}, \bibinfo {author} {\bibfnamefont
		{J.}~\bibnamefont {Wosnitza}},  \emph {et~al.},\ }\href@noop {} {\bibfield
	{journal} {\bibinfo  {journal} {Physical Review B}\ }\textbf {\bibinfo
		{volume} {100}},\ \bibinfo {pages} {224417} (\bibinfo {year}
	{2019}{\natexlab{b}})}\BibitemShut {NoStop}%
	\bibitem [{\citenamefont {Sala}\ \emph {et~al.}(2019)\citenamefont {Sala},
		\citenamefont {Stone}, \citenamefont {Rai}, \citenamefont {May},
		\citenamefont {Parker}, \citenamefont {Hal{\'a}sz}, \citenamefont {Cheng},
		\citenamefont {Ehlers}, \citenamefont {Garlea}, \citenamefont {Zhang} \emph
		{et~al.}}]{sala2019crystal}%
	\BibitemOpen
	\bibfield  {author} {\bibinfo {author} {\bibfnamefont {G.}~\bibnamefont
			{Sala}}, \bibinfo {author} {\bibfnamefont {M.}~\bibnamefont {Stone}},
		\bibinfo {author} {\bibfnamefont {B.~K.}\ \bibnamefont {Rai}}, \bibinfo
		{author} {\bibfnamefont {A.~F.}\ \bibnamefont {May}}, \bibinfo {author}
		{\bibfnamefont {D.~S.}\ \bibnamefont {Parker}}, \bibinfo {author}
		{\bibfnamefont {G.~B.}\ \bibnamefont {Hal{\'a}sz}}, \bibinfo {author}
		{\bibfnamefont {Y.~Q.}\ \bibnamefont {Cheng}}, \bibinfo {author}
		{\bibfnamefont {G.}~\bibnamefont {Ehlers}}, \bibinfo {author} {\bibfnamefont
			{V.~O.}\ \bibnamefont {Garlea}}, \bibinfo {author} {\bibfnamefont
			{Q.}~\bibnamefont {Zhang}},  \emph {et~al.},\ }\href@noop {} {\bibfield
		{journal} {\bibinfo  {journal} {arXiv preprint arXiv:1907.10627}\ } (\bibinfo
		{year} {2019})}\BibitemShut {NoStop}%
	\bibitem [{\citenamefont {Rau}\ \emph {et~al.}(2016)\citenamefont {Rau},
	\citenamefont {Wu}, \citenamefont {May}, \citenamefont {Poudel},
	\citenamefont {Winn}, \citenamefont {Garlea}, \citenamefont {Huq},
	\citenamefont {Whitfield}, \citenamefont {Taylor}, \citenamefont {Lumsden}, \citenamefont {Gingras}, \citenamefont {Christianson}
}]{rau2016anisotropic}%
\BibitemOpen
\bibfield  {author} {\bibinfo {author} {\bibfnamefont {J.~G.}~\bibnamefont
		{Rau}}, \bibinfo {author} {\bibfnamefont {L.~S.}~\bibnamefont {Wu}}, \bibinfo
	{author} {\bibfnamefont {A.~F.}~\bibnamefont {May}}, \bibinfo {author}
	{\bibfnamefont {L.}~\bibnamefont {Poudel}}, \bibinfo {author} {\bibfnamefont
		{B.}~\bibnamefont {Winn}}, \bibinfo {author} {\bibfnamefont {V.~O.}~\bibnamefont
		{Garlea}}, \bibinfo {author} {\bibfnamefont {A.}~\bibnamefont {Huq}},
	\bibinfo {author} {\bibfnamefont {P.}~\bibnamefont {Whitfield}}, \bibinfo
	{author} {\bibfnamefont {A.~E.}~\bibnamefont {Taylor}}, \bibinfo {author}
	{\bibfnamefont {M.~D.}~\bibnamefont {Lumsden}}, \bibinfo {author}
	{\bibfnamefont {M.~J.~P.}~\bibnamefont {Gingras}}, \bibinfo {author}
	{\bibfnamefont {A.~D.}~\bibnamefont {Christianson}},\ }\href@noop {}
{\bibfield  {journal} {\bibinfo  {journal} {Physical Review Letters}\
	}\textbf {\bibinfo {volume} {116}},\ \bibinfo {pages} {257204} (\bibinfo
	{year} {2016})}\BibitemShut {NoStop}%
	\bibitem [{\citenamefont {Sanjeewa}\ \emph {et~al.}(2018)\citenamefont
	{Sanjeewa}, \citenamefont {Ross}, \citenamefont {Sarkis}, \citenamefont
	{Nair}, \citenamefont {McMillen},\ and\ \citenamefont
	{Kolis}}]{sanjeewa2018single}%
\BibitemOpen
\bibfield  {author} {\bibinfo {author} {\bibfnamefont {L.~D.}\ \bibnamefont
		{Sanjeewa}}, \bibinfo {author} {\bibfnamefont {K.~A.}\ \bibnamefont {Ross}},
	\bibinfo {author} {\bibfnamefont {C.~L.}\ \bibnamefont {Sarkis}}, \bibinfo
	{author} {\bibfnamefont {H.~S.}\ \bibnamefont {Nair}}, \bibinfo {author}
	{\bibfnamefont {C.~D.}\ \bibnamefont {McMillen}}, \ and\ \bibinfo {author}
	{\bibfnamefont {J.~W.}\ \bibnamefont {Kolis}},\ }\href@noop {} {\bibfield
	{journal} {\bibinfo  {journal} {Inorganic Chemistry}\ }\textbf {\bibinfo
		{volume} {57}},\ \bibinfo {pages} {12456} (\bibinfo {year}
	{2018})}\BibitemShut {NoStop}%
\bibitem [{\citenamefont {Dun}\ \emph {et~al.}(2015)\citenamefont {Dun},
	\citenamefont {Li}, \citenamefont {Freitas}, \citenamefont {Arrighi},
	\citenamefont {DelaCruz}, \citenamefont {Lee}, \citenamefont {Choi},
	\citenamefont {Cao}, \citenamefont {Silverstein}, \citenamefont {Wiebe} \emph
	{et~al.}}]{dun2015antiferromagnetic}%
\BibitemOpen
\bibfield  {author} {\bibinfo {author} {\bibfnamefont {Z.~L.}\ \bibnamefont
		{Dun}}, \bibinfo {author} {\bibfnamefont {X.}~\bibnamefont {Li}}, \bibinfo
	{author} {\bibfnamefont {R.~S.}\ \bibnamefont {Freitas}}, \bibinfo {author}
	{\bibfnamefont {E.}~\bibnamefont {Arrighi}}, \bibinfo {author} {\bibfnamefont
		{C.~R.}\ \bibnamefont {DelaCruz}}, \bibinfo {author} {\bibfnamefont
		{M.}~\bibnamefont {Lee}}, \bibinfo {author} {\bibfnamefont {E.~S.}\
		\bibnamefont {Choi}}, \bibinfo {author} {\bibfnamefont {H.~B.}\ \bibnamefont
		{Cao}}, \bibinfo {author} {\bibfnamefont {H.~J.}\ \bibnamefont
		{Silverstein}}, \bibinfo {author} {\bibfnamefont {C.~R.}\ \bibnamefont
		{Wiebe}},  \emph {et~al.},\ }\href@noop {} {\bibfield  {journal} {\bibinfo
		{journal} {Physical Review B}\ }\textbf {\bibinfo {volume} {92}},\ \bibinfo
	{pages} {140407(R)} (\bibinfo {year} {2015})}\BibitemShut {NoStop}%
	\bibitem [{\citenamefont {Hallas}\ \emph {et~al.}(2018)\citenamefont {Hallas},
		\citenamefont {Gaudet},\ and\ \citenamefont
		{Gaulin}}]{hallas2018experimental}%
	\BibitemOpen
	\bibfield  {author} {\bibinfo {author} {\bibfnamefont {A.~M.}\ \bibnamefont
			{Hallas}}, \bibinfo {author} {\bibfnamefont {J.}~\bibnamefont {Gaudet}}, \
		and\ \bibinfo {author} {\bibfnamefont {B.~D.}\ \bibnamefont {Gaulin}},\
	}\href@noop {} {\bibfield  {journal} {\bibinfo  {journal} {Annual Review of
				Condensed Matter Physics}\ }\textbf {\bibinfo {volume} {9}},\ \bibinfo
		{pages} {105} (\bibinfo {year} {2018})}\BibitemShut {NoStop}%
	\bibitem [{\citenamefont {Rau}\ and\ \citenamefont
		{Gingras}(2019)}]{rau2019frustrated}%
	\BibitemOpen
	\bibfield  {author} {\bibinfo {author} {\bibfnamefont {J.~G.}\ \bibnamefont
			{Rau}}\ and\ \bibinfo {author} {\bibfnamefont {M.~J.~P.}\ \bibnamefont
			{Gingras}},\ }\href@noop {} {\bibfield  {journal} {\bibinfo  {journal}
			{Annual Review of Condensed Matter Physics}\ }\textbf {\bibinfo {volume}
			{10}},\ \bibinfo {pages} {357} (\bibinfo {year} {2019})}\BibitemShut
	{NoStop}%
	\bibitem [{\citenamefont {Ross}\ \emph {et~al.}(2011)\citenamefont {Ross},
		\citenamefont {Savary}, \citenamefont {Gaulin},\ and\ \citenamefont
		{Balents}}]{ross2011quantum}%
	\BibitemOpen
	\bibfield  {author} {\bibinfo {author} {\bibfnamefont {K.~A.}\ \bibnamefont
			{Ross}}, \bibinfo {author} {\bibfnamefont {L.}~\bibnamefont {Savary}},
		\bibinfo {author} {\bibfnamefont {B.~D.}\ \bibnamefont {Gaulin}}, \ and\
		\bibinfo {author} {\bibfnamefont {L.}~\bibnamefont {Balents}},\ }\href@noop
	{} {\bibfield  {journal} {\bibinfo  {journal} {Physical Review X}\ }\textbf
		{\bibinfo {volume} {1}},\ \bibinfo {pages} {021002} (\bibinfo {year}
		{2011})}\BibitemShut {NoStop}%
	\bibitem [{\citenamefont {Savary}\ \emph {et~al.}(2012)\citenamefont {Savary},
		\citenamefont {Ross}, \citenamefont {Gaulin}, \citenamefont {Ruff},\ and\
		\citenamefont {Balents}}]{savary2012order}%
	\BibitemOpen
	\bibfield  {author} {\bibinfo {author} {\bibfnamefont {L.}~\bibnamefont
			{Savary}}, \bibinfo {author} {\bibfnamefont {K.~A.}\ \bibnamefont {Ross}},
		\bibinfo {author} {\bibfnamefont {B.~D.}\ \bibnamefont {Gaulin}}, \bibinfo
		{author} {\bibfnamefont {J.~P.~C.}\ \bibnamefont {Ruff}}, \ and\ \bibinfo
		{author} {\bibfnamefont {L.}~\bibnamefont {Balents}},\ }\href@noop {}
	{\bibfield  {journal} {\bibinfo  {journal} {Physical Review Letters}\
		}\textbf {\bibinfo {volume} {109}},\ \bibinfo {pages} {167201} (\bibinfo
		{year} {2012})}\BibitemShut {NoStop}%
	\bibitem [{\citenamefont {Robert}\ \emph {et~al.}(2015)\citenamefont {Robert},
		\citenamefont {Lhotel}, \citenamefont {Remenyi}, \citenamefont {Sahling},
		\citenamefont {Mirebeau}, \citenamefont {Decorse}, \citenamefont {Canals},\
		and\ \citenamefont {Petit}}]{robert2015spin}%
	\BibitemOpen
	\bibfield  {author} {\bibinfo {author} {\bibfnamefont {J.}~\bibnamefont
			{Robert}}, \bibinfo {author} {\bibfnamefont {E.}~\bibnamefont {Lhotel}},
		\bibinfo {author} {\bibfnamefont {G.}~\bibnamefont {Remenyi}}, \bibinfo
		{author} {\bibfnamefont {S.}~\bibnamefont {Sahling}}, \bibinfo {author}
		{\bibfnamefont {I.}~\bibnamefont {Mirebeau}}, \bibinfo {author}
		{\bibfnamefont {C.}~\bibnamefont {Decorse}}, \bibinfo {author} {\bibfnamefont
			{B.}~\bibnamefont {Canals}}, \ and\ \bibinfo {author} {\bibfnamefont
			{S.}~\bibnamefont {Petit}},\ }\href@noop {} {\bibfield  {journal} {\bibinfo
			{journal} {Physical Review B}\ }\textbf {\bibinfo {volume} {92}},\ \bibinfo
		{pages} {064425} (\bibinfo {year} {2015})}\BibitemShut {NoStop}%
	\bibitem [{\citenamefont {Thompson}\ \emph {et~al.}(2017)\citenamefont
		{Thompson}, \citenamefont {McClarty}, \citenamefont {Prabhakaran},
		\citenamefont {Cabrera}, \citenamefont {Guidi},\ and\ \citenamefont
		{Coldea}}]{thompson2017quasiparticle}%
	\BibitemOpen
	\bibfield  {author} {\bibinfo {author} {\bibfnamefont {J.~D.}\ \bibnamefont
			{Thompson}}, \bibinfo {author} {\bibfnamefont {P.~A.}\ \bibnamefont
			{McClarty}}, \bibinfo {author} {\bibfnamefont {D.}~\bibnamefont
			{Prabhakaran}}, \bibinfo {author} {\bibfnamefont {I.}~\bibnamefont
			{Cabrera}}, \bibinfo {author} {\bibfnamefont {T.}~\bibnamefont {Guidi}}, \
		and\ \bibinfo {author} {\bibfnamefont {R.}~\bibnamefont {Coldea}},\
	}\href@noop {} {\bibfield  {journal} {\bibinfo  {journal} {Physical Review
				Letters}\ }\textbf {\bibinfo {volume} {119}},\ \bibinfo {pages} {057203}
		(\bibinfo {year} {2017})}\BibitemShut {NoStop}%
	\bibitem [{\citenamefont {Zhitomirsky}\ \emph {et~al.}(2012)\citenamefont
		{Zhitomirsky}, \citenamefont {Gvozdikova}, \citenamefont {Holdsworth},\ and\
		\citenamefont {Moessner}}]{zhitomirsky2012quantum}%
	\BibitemOpen
	\bibfield  {author} {\bibinfo {author} {\bibfnamefont {M.~E.}\ \bibnamefont
			{Zhitomirsky}}, \bibinfo {author} {\bibfnamefont {M.~V.}\ \bibnamefont
			{Gvozdikova}}, \bibinfo {author} {\bibfnamefont {P.~C.~W.}\ \bibnamefont
			{Holdsworth}}, \ and\ \bibinfo {author} {\bibfnamefont {R.}~\bibnamefont
			{Moessner}},\ }\href@noop {} {\bibfield  {journal} {\bibinfo  {journal}
			{Physical Review Letters}\ }\textbf {\bibinfo {volume} {109}},\ \bibinfo
		{pages} {077204} (\bibinfo {year} {2012})}\BibitemShut {NoStop}%
	\bibitem [{\citenamefont {Curnoe}(2008)}]{curnoe2008structural}%
	\BibitemOpen
	\bibfield  {author} {\bibinfo {author} {\bibfnamefont {S.~H.}\ \bibnamefont
			{Curnoe}},\ }\href@noop {} {\bibfield  {journal} {\bibinfo  {journal}
			{Physical Review B}\ }\textbf {\bibinfo {volume} {78}},\ \bibinfo {pages}
		{094418} (\bibinfo {year} {2008})}\BibitemShut {NoStop}%
	\bibitem [{\citenamefont {Wong}\ \emph {et~al.}(2013)\citenamefont {Wong},
		\citenamefont {Hao},\ and\ \citenamefont {Gingras}}]{WongPRB2013}%
	\BibitemOpen
	\bibfield  {author} {\bibinfo {author} {\bibfnamefont {A.~W.~C.}\
			\bibnamefont {Wong}}, \bibinfo {author} {\bibfnamefont {Z.}~\bibnamefont
			{Hao}}, \ and\ \bibinfo {author} {\bibfnamefont {M.~J.~P.}\ \bibnamefont
			{Gingras}},\ }\href@noop {} {\bibfield  {journal} {\bibinfo  {journal}
			{Physical Review B}\ }\textbf {\bibinfo {volume} {88}},\ \bibinfo {pages}
		{144402} (\bibinfo {year} {2013})}\BibitemShut {NoStop}%
	\bibitem [{\citenamefont {Yan}\ \emph {et~al.}(2017)\citenamefont {Yan},
		\citenamefont {Benton}, \citenamefont {Jaubert},\ and\ \citenamefont
		{Shannon}}]{yan2017theory}%
	\BibitemOpen
	\bibfield  {author} {\bibinfo {author} {\bibfnamefont {H.}~\bibnamefont
			{Yan}}, \bibinfo {author} {\bibfnamefont {O.}~\bibnamefont {Benton}},
		\bibinfo {author} {\bibfnamefont {L.}~\bibnamefont {Jaubert}}, \ and\
		\bibinfo {author} {\bibfnamefont {N.}~\bibnamefont {Shannon}},\ }\href@noop
	{} {\bibfield  {journal} {\bibinfo  {journal} {Physical Review B}\ }\textbf
		{\bibinfo {volume} {95}},\ \bibinfo {pages} {094422} (\bibinfo {year}
		{2017})}\BibitemShut {NoStop}%
	\bibitem [{\citenamefont {Yasui}\ \emph {et~al.}(2003)\citenamefont {Yasui},
		\citenamefont {Soda}, \citenamefont {Iikubo}, \citenamefont {Ito},
		\citenamefont {Sato}, \citenamefont {Hamaguchi}, \citenamefont {Matsushita},
		\citenamefont {Wada}, \citenamefont {Takeuchi}, \citenamefont {Aso} \emph
		{et~al.}}]{yasui2003ferromagnetic}%
	\BibitemOpen
	\bibfield  {author} {\bibinfo {author} {\bibfnamefont {Y.}~\bibnamefont
			{Yasui}}, \bibinfo {author} {\bibfnamefont {M.}~\bibnamefont {Soda}},
		\bibinfo {author} {\bibfnamefont {S.}~\bibnamefont {Iikubo}}, \bibinfo
		{author} {\bibfnamefont {M.}~\bibnamefont {Ito}}, \bibinfo {author}
		{\bibfnamefont {M.}~\bibnamefont {Sato}}, \bibinfo {author} {\bibfnamefont
			{N.}~\bibnamefont {Hamaguchi}}, \bibinfo {author} {\bibfnamefont
			{T.}~\bibnamefont {Matsushita}}, \bibinfo {author} {\bibfnamefont
			{N.}~\bibnamefont {Wada}}, \bibinfo {author} {\bibfnamefont {T.}~\bibnamefont
			{Takeuchi}}, \bibinfo {author} {\bibfnamefont {N.}~\bibnamefont {Aso}},
		\emph {et~al.},\ }\href@noop {} {\bibfield  {journal} {\bibinfo  {journal}
			{Journal of the Physical Society of Japan}\ }\textbf {\bibinfo {volume}
			{72}},\ \bibinfo {pages} {3014} (\bibinfo {year} {2003})}\BibitemShut
	{NoStop}%
	\bibitem [{\citenamefont {Gaudet}\ \emph {et~al.}(2016)\citenamefont {Gaudet},
		\citenamefont {Ross}, \citenamefont {Kermarrec}, \citenamefont {Butch},
		\citenamefont {Ehlers}, \citenamefont {Dabkowska},\ and\ \citenamefont
		{Gaulin}}]{gaudet2016gapless}%
	\BibitemOpen
	\bibfield  {author} {\bibinfo {author} {\bibfnamefont {J.}~\bibnamefont
			{Gaudet}}, \bibinfo {author} {\bibfnamefont {K.~A.}\ \bibnamefont {Ross}},
		\bibinfo {author} {\bibfnamefont {E.}~\bibnamefont {Kermarrec}}, \bibinfo
		{author} {\bibfnamefont {N.~P.}\ \bibnamefont {Butch}}, \bibinfo {author}
		{\bibfnamefont {G.}~\bibnamefont {Ehlers}}, \bibinfo {author} {\bibfnamefont
			{H.~A.}\ \bibnamefont {Dabkowska}}, \ and\ \bibinfo {author} {\bibfnamefont
			{B.~D.}\ \bibnamefont {Gaulin}},\ }\href@noop {} {\bibfield  {journal}
		{\bibinfo  {journal} {Physical Review B}\ }\textbf {\bibinfo {volume} {93}},\
		\bibinfo {pages} {064406} (\bibinfo {year} {2016})}\BibitemShut {NoStop}%
	\bibitem [{\citenamefont {Yaouanc}\ \emph {et~al.}(2016)\citenamefont
		{Yaouanc}, \citenamefont {de~R{\'e}otier}, \citenamefont {Keller},
		\citenamefont {Roessli},\ and\ \citenamefont {Forget}}]{yaouanc2016novel}%
	\BibitemOpen
	\bibfield  {author} {\bibinfo {author} {\bibfnamefont {A.}~\bibnamefont
			{Yaouanc}}, \bibinfo {author} {\bibfnamefont {P.~D.}\ \bibnamefont
			{de~R{\'e}otier}}, \bibinfo {author} {\bibfnamefont {L.}~\bibnamefont
			{Keller}}, \bibinfo {author} {\bibfnamefont {B.}~\bibnamefont {Roessli}}, \
		and\ \bibinfo {author} {\bibfnamefont {A.}~\bibnamefont {Forget}},\
	}\href@noop {} {\bibfield  {journal} {\bibinfo  {journal} {Journal of
				Physics: Condensed Matter}\ }\textbf {\bibinfo {volume} {28}},\ \bibinfo
		{pages} {426002} (\bibinfo {year} {2016})}\BibitemShut {NoStop}%
	\bibitem [{\citenamefont {Yaouanc}\ \emph {et~al.}(2013)\citenamefont
		{Yaouanc}, \citenamefont {De~R{\'e}otier}, \citenamefont {Bonville},
		\citenamefont {Hodges}, \citenamefont {Glazkov}, \citenamefont {Keller},
		\citenamefont {Sikolenko}, \citenamefont {Bartkowiak}, \citenamefont {Amato},
		\citenamefont {Baines} \emph {et~al.}}]{yaouanc2013dynamical}%
	\BibitemOpen
	\bibfield  {author} {\bibinfo {author} {\bibfnamefont {A.}~\bibnamefont
			{Yaouanc}}, \bibinfo {author} {\bibfnamefont {P.~D.}\ \bibnamefont
			{De~R{\'e}otier}}, \bibinfo {author} {\bibfnamefont {P.}~\bibnamefont
			{Bonville}}, \bibinfo {author} {\bibfnamefont {J.~A.}\ \bibnamefont
			{Hodges}}, \bibinfo {author} {\bibfnamefont {V.}~\bibnamefont {Glazkov}},
		\bibinfo {author} {\bibfnamefont {L.}~\bibnamefont {Keller}}, \bibinfo
		{author} {\bibfnamefont {V.}~\bibnamefont {Sikolenko}}, \bibinfo {author}
		{\bibfnamefont {M.}~\bibnamefont {Bartkowiak}}, \bibinfo {author}
		{\bibfnamefont {A.}~\bibnamefont {Amato}}, \bibinfo {author} {\bibfnamefont
			{C.}~\bibnamefont {Baines}},  \emph {et~al.},\ }\href@noop {} {\bibfield
		{journal} {\bibinfo  {journal} {Physical Review Letters}\ }\textbf {\bibinfo
			{volume} {110}},\ \bibinfo {pages} {127207} (\bibinfo {year}
		{2013})}\BibitemShut {NoStop}%
	\bibitem [{\citenamefont {Lago}\ \emph {et~al.}(2014)\citenamefont {Lago},
	\citenamefont {{\v{Z}}ivkovi{\'c}}, \citenamefont {Piatek}, \citenamefont
	{{\'A}lvarez}, \citenamefont {H{\"u}vonen}, \citenamefont {Pratt},
	\citenamefont {D{\'\i}az},\ and\ \citenamefont {Rojo}}]{lago2014glassy}%
\BibitemOpen
\bibfield  {author} {\bibinfo {author} {\bibfnamefont {J.}~\bibnamefont
		{Lago}}, \bibinfo {author} {\bibfnamefont {I.}~\bibnamefont
		{{\v{Z}}ivkovi{\'c}}}, \bibinfo {author} {\bibfnamefont {J.~O.}~\bibnamefont
		{Piatek}}, \bibinfo {author} {\bibfnamefont {P.}~\bibnamefont {{\'A}lvarez}},
	\bibinfo {author} {\bibfnamefont {D.}~\bibnamefont {H{\"u}vonen}}, \bibinfo
	{author} {\bibfnamefont {F.~L.}\ \bibnamefont {Pratt}}, \bibinfo {author}
	{\bibfnamefont {M.}~\bibnamefont {D{\'\i}az}}, \ and\ \bibinfo {author}
	{\bibfnamefont {T.}~\bibnamefont {Rojo}},\ }\href@noop {} {\bibfield
	{journal} {\bibinfo  {journal} {Physical Review B}\ }\textbf {\bibinfo
		{volume} {89}},\ \bibinfo {pages} {024421} (\bibinfo {year}
	{2014})}\BibitemShut {NoStop}%
	\bibitem [{\citenamefont {Rau}\ \emph {et~al.}(2019)\citenamefont {Rau},
		\citenamefont {Moessner},\ and\ \citenamefont {McClarty}}]{rau2019magnon}%
	\BibitemOpen
	\bibfield  {author} {\bibinfo {author} {\bibfnamefont {J.~G.}\ \bibnamefont
			{Rau}}, \bibinfo {author} {\bibfnamefont {R.}~\bibnamefont {Moessner}}, \
		and\ \bibinfo {author} {\bibfnamefont {P.~A.}\ \bibnamefont {McClarty}},\
	}\href@noop {} {\bibfield  {journal} {\bibinfo  {journal} {Physical Review
				B}\ }\textbf {\bibinfo {volume} {100}},\ \bibinfo {pages} {104423} (\bibinfo
		{year} {2019})}\BibitemShut {NoStop}%
	\bibitem [{\citenamefont {Scheie}\ \emph {et~al.}(2019)\citenamefont {Scheie},
		\citenamefont {Kindervater}, \citenamefont {Zhang}, \citenamefont
		{Changlani}, \citenamefont {Sala}, \citenamefont {Ehlers}, \citenamefont
		{Heinemann}, \citenamefont {Tucker}, \citenamefont {Koohpayeh},\ and\
		\citenamefont {Broholm}}]{scheie2019multiphase}%
	\BibitemOpen
	\bibfield  {author} {\bibinfo {author} {\bibfnamefont {A.}~\bibnamefont
			{Scheie}}, \bibinfo {author} {\bibfnamefont {J.}~\bibnamefont {Kindervater}},
		\bibinfo {author} {\bibfnamefont {S.}~\bibnamefont {Zhang}}, \bibinfo
		{author} {\bibfnamefont {H.}~\bibnamefont {Changlani}}, \bibinfo {author}
		{\bibfnamefont {G.}~\bibnamefont {Sala}}, \bibinfo {author} {\bibfnamefont
			{G.}~\bibnamefont {Ehlers}}, \bibinfo {author} {\bibfnamefont
			{A.}~\bibnamefont {Heinemann}}, \bibinfo {author} {\bibfnamefont
			{G.}~\bibnamefont {Tucker}}, \bibinfo {author} {\bibfnamefont
			{S.}~\bibnamefont {Koohpayeh}}, \ and\ \bibinfo {author} {\bibfnamefont
			{C.}~\bibnamefont {Broholm}},\ }\href@noop {} {\bibfield  {journal} {\bibinfo
			{journal} {arXiv preprint arXiv:1912.04913}\ } (\bibinfo {year}
		{2019})}\BibitemShut {NoStop}%
	\bibitem [{\citenamefont {Dem'yanets}\ \emph {et~al.}(1988)\citenamefont
		{Dem'yanets}, \citenamefont {Radaev}, \citenamefont {Mamin},\ and\
		\citenamefont {Maksimov}}]{dem1988synthesis}%
	\BibitemOpen
	\bibfield  {author} {\bibinfo {author} {\bibfnamefont {L.~N.}\ \bibnamefont
			{Dem'yanets}}, \bibinfo {author} {\bibfnamefont {S.~F.}\ \bibnamefont
			{Radaev}}, \bibinfo {author} {\bibfnamefont {B.~F.}\ \bibnamefont {Mamin}}, \
		and\ \bibinfo {author} {\bibfnamefont {B.~A.}\ \bibnamefont {Maksimov}},\
	}\href@noop {} {\bibfield  {journal} {\bibinfo  {journal} {Journal of
				Structural Chemistry}\ }\textbf {\bibinfo {volume} {29}},\ \bibinfo {pages}
		{485} (\bibinfo {year} {1988})}\BibitemShut {NoStop}%
	\bibitem [{\citenamefont {Becker}\ and\ \citenamefont
		{Felsche}(1987)}]{becker1987phases}%
	\BibitemOpen
	\bibfield  {author} {\bibinfo {author} {\bibfnamefont {U.~W.}\ \bibnamefont
			{Becker}}\ and\ \bibinfo {author} {\bibfnamefont {J.}~\bibnamefont
			{Felsche}},\ }\href@noop {} {\bibfield  {journal} {\bibinfo  {journal}
			{Journal of the Less Common Metals}\ }\textbf {\bibinfo {volume} {128}},\
		\bibinfo {pages} {269} (\bibinfo {year} {1987})}\BibitemShut {NoStop}%
	\bibitem [{\citenamefont {Cai}\ \emph {et~al.}(2011)\citenamefont {Cai},
		\citenamefont {Arias},\ and\ \citenamefont {Nino}}]{cai2011tolerance}%
	\BibitemOpen
	\bibfield  {author} {\bibinfo {author} {\bibfnamefont {L.}~\bibnamefont
			{Cai}}, \bibinfo {author} {\bibfnamefont {A.~L.}\ \bibnamefont {Arias}}, \
		and\ \bibinfo {author} {\bibfnamefont {J.~C.}\ \bibnamefont {Nino}},\
	}\href@noop {} {\bibfield  {journal} {\bibinfo  {journal} {Journal of
				Materials Chemistry}\ }\textbf {\bibinfo {volume} {21}},\ \bibinfo {pages}
		{3611} (\bibinfo {year} {2011})}\BibitemShut {NoStop}%
	\bibitem [{\citenamefont {Hallas}\ \emph
		{et~al.}(2016{\natexlab{a}})\citenamefont {Hallas}, \citenamefont {Gaudet},
		\citenamefont {Wilson}, \citenamefont {Munsie}, \citenamefont {Aczel},
		\citenamefont {Stone}, \citenamefont {Freitas}, \citenamefont
		{Arevalo-Lopez}, \citenamefont {Attfield}, \citenamefont {Tachibana} \emph
		{et~al.}}]{hallas2016xy}%
	\BibitemOpen
	\bibfield  {author} {\bibinfo {author} {\bibfnamefont {A.~M.}\ \bibnamefont
			{Hallas}}, \bibinfo {author} {\bibfnamefont {J.}~\bibnamefont {Gaudet}},
		\bibinfo {author} {\bibfnamefont {M.~N.}\ \bibnamefont {Wilson}}, \bibinfo
		{author} {\bibfnamefont {T.~J.}\ \bibnamefont {Munsie}}, \bibinfo {author}
		{\bibfnamefont {A.~A.}\ \bibnamefont {Aczel}}, \bibinfo {author}
		{\bibfnamefont {M.~B.}\ \bibnamefont {Stone}}, \bibinfo {author}
		{\bibfnamefont {R.~S.}\ \bibnamefont {Freitas}}, \bibinfo {author}
		{\bibfnamefont {A.~M.}\ \bibnamefont {Arevalo-Lopez}}, \bibinfo {author}
		{\bibfnamefont {J.~P.}\ \bibnamefont {Attfield}}, \bibinfo {author}
		{\bibfnamefont {M.}~\bibnamefont {Tachibana}},  \emph {et~al.},\ }\href@noop
	{} {\bibfield  {journal} {\bibinfo  {journal} {Physical Review B}\ }\textbf
		{\bibinfo {volume} {93}},\ \bibinfo {pages} {104405} (\bibinfo {year}
		{2016}{\natexlab{a}})}\BibitemShut {NoStop}%
	\bibitem [{\citenamefont {Shannon}\ and\ \citenamefont
		{Sleight}(1968)}]{shannon1968synthesis}%
	\BibitemOpen
	\bibfield  {author} {\bibinfo {author} {\bibfnamefont {R.~D.}\ \bibnamefont
			{Shannon}}\ and\ \bibinfo {author} {\bibfnamefont {A.~W.}\ \bibnamefont
			{Sleight}},\ }\href@noop {} {\bibfield  {journal} {\bibinfo  {journal}
			{Inorganic Chemistry}\ }\textbf {\bibinfo {volume} {7}},\ \bibinfo {pages}
		{1649} (\bibinfo {year} {1968})}\BibitemShut {NoStop}%
\bibitem [{\citenamefont {Antlauf}\ \emph {et~al.}(2019)\citenamefont
	{Antlauf}, \citenamefont {Taniguchi}, \citenamefont {Wagler}, \citenamefont
	{Schwarz},\ and\ \citenamefont {Kroke}}]{antlauf2019synthesis}%
\BibitemOpen
\bibfield  {author} {\bibinfo {author} {\bibfnamefont {M.}~\bibnamefont
		{Antlauf}}, \bibinfo {author} {\bibfnamefont {T.}~\bibnamefont {Taniguchi}},
	\bibinfo {author} {\bibfnamefont {J.}~\bibnamefont {Wagler}}, \bibinfo
	{author} {\bibfnamefont {M.~R.}\ \bibnamefont {Schwarz}}, \ and\ \bibinfo
	{author} {\bibfnamefont {E.}~\bibnamefont {Kroke}},\ }\href@noop {}
{\bibfield  {journal} {\bibinfo  {journal} {Crystal Growth \& Design}\
	}\textbf {\bibinfo {volume} {19}},\ \bibinfo {pages} {5538} (\bibinfo {year}
	{2019})}\BibitemShut {NoStop}%
	\bibitem [{\citenamefont {Dun}\ \emph {et~al.}(2014)\citenamefont {Dun},
		\citenamefont {Lee}, \citenamefont {Choi}, \citenamefont {Hallas},
		\citenamefont {Wiebe}, \citenamefont {Gardner}, \citenamefont {Arrighi},
		\citenamefont {Freitas}, \citenamefont {Arevalo-Lopez}, \citenamefont
		{Attfield} \emph {et~al.}}]{dun2014chemical}%
	\BibitemOpen
	\bibfield  {author} {\bibinfo {author} {\bibfnamefont {Z.~L.}\ \bibnamefont
			{Dun}}, \bibinfo {author} {\bibfnamefont {M.}~\bibnamefont {Lee}}, \bibinfo
		{author} {\bibfnamefont {E.~S.}\ \bibnamefont {Choi}}, \bibinfo {author}
		{\bibfnamefont {A.~M.}\ \bibnamefont {Hallas}}, \bibinfo {author}
		{\bibfnamefont {C.~R.}\ \bibnamefont {Wiebe}}, \bibinfo {author}
		{\bibfnamefont {J.~S.}\ \bibnamefont {Gardner}}, \bibinfo {author}
		{\bibfnamefont {E.}~\bibnamefont {Arrighi}}, \bibinfo {author} {\bibfnamefont
			{R.~S.}\ \bibnamefont {Freitas}}, \bibinfo {author} {\bibfnamefont {A.~M.}\
			\bibnamefont {Arevalo-Lopez}}, \bibinfo {author} {\bibfnamefont {J.~P.}\
			\bibnamefont {Attfield}},  \emph {et~al.},\ }\href@noop {} {\bibfield
		{journal} {\bibinfo  {journal} {Physical Review B}\ }\textbf {\bibinfo
			{volume} {89}},\ \bibinfo {pages} {064401} (\bibinfo {year}
		{2014})}\BibitemShut {NoStop}%
	\bibitem [{\citenamefont {Hallas}\ \emph
		{et~al.}(2016{\natexlab{b}})\citenamefont {Hallas}, \citenamefont {Gaudet},
		\citenamefont {Butch}, \citenamefont {Tachibana}, \citenamefont {Freitas},
		\citenamefont {Luke}, \citenamefont {Wiebe},\ and\ \citenamefont
		{Gaulin}}]{hallas2016universal}%
	\BibitemOpen
	\bibfield  {author} {\bibinfo {author} {\bibfnamefont {A.~M.}\ \bibnamefont
			{Hallas}}, \bibinfo {author} {\bibfnamefont {J.}~\bibnamefont {Gaudet}},
		\bibinfo {author} {\bibfnamefont {N.~P.}\ \bibnamefont {Butch}}, \bibinfo
		{author} {\bibfnamefont {M.}~\bibnamefont {Tachibana}}, \bibinfo {author}
		{\bibfnamefont {R.~S.}\ \bibnamefont {Freitas}}, \bibinfo {author}
		{\bibfnamefont {G.~M.}\ \bibnamefont {Luke}}, \bibinfo {author}
		{\bibfnamefont {C.~R.}\ \bibnamefont {Wiebe}}, \ and\ \bibinfo {author}
		{\bibfnamefont {B.~D.}\ \bibnamefont {Gaulin}},\ }\href@noop {} {\bibfield
		{journal} {\bibinfo  {journal} {Physical Review B}\ }\textbf {\bibinfo
			{volume} {93}},\ \bibinfo {pages} {100403(R)} (\bibinfo {year}
		{2016}{\natexlab{b}})}\BibitemShut {NoStop}%
	\bibitem [{\citenamefont {Applegate}\ \emph {et~al.}(2012)\citenamefont
		{Applegate}, \citenamefont {Hayre}, \citenamefont {Singh}, \citenamefont
		{Lin}, \citenamefont {Day},\ and\ \citenamefont {Gingras}}]{Applegate2012a}%
	\BibitemOpen
	\bibfield  {author} {\bibinfo {author} {\bibfnamefont {R.}~\bibnamefont
			{Applegate}}, \bibinfo {author} {\bibfnamefont {N.~R.}\ \bibnamefont
			{Hayre}}, \bibinfo {author} {\bibfnamefont {R.~R.~P.}\ \bibnamefont {Singh}},
		\bibinfo {author} {\bibfnamefont {T.}~\bibnamefont {Lin}}, \bibinfo {author}
		{\bibfnamefont {A.~G.~R.}\ \bibnamefont {Day}}, \ and\ \bibinfo {author}
		{\bibfnamefont {M.~J.~P.}\ \bibnamefont {Gingras}},\ }\href@noop {}
	{\bibfield  {journal} {\bibinfo  {journal} {Physical Review Letters}\
		}\textbf {\bibinfo {volume} {109}},\ \bibinfo {pages} {097205} (\bibinfo
		{year} {2012})}\BibitemShut {NoStop}%
	\bibitem [{\citenamefont {Hayre}\ \emph {et~al.}(2013)\citenamefont {Hayre},
		\citenamefont {Ross}, \citenamefont {Applegate}, \citenamefont {Lin},
		 \citenamefont {Singh}, \citenamefont {Gaulin},\ and\ \citenamefont
		{Gingras}}]{Hayre2012}%
	\BibitemOpen
	\bibfield  {author} {\bibinfo {author} {\bibfnamefont {N.~R.}\ \bibnamefont
			{Hayre}}, \bibinfo {author} {\bibfnamefont {K.~A.}\ \bibnamefont {Ross}},
		\bibinfo {author} {\bibfnamefont {R.}~\bibnamefont {Applegate}}, \bibinfo
		{author} {\bibfnamefont {T.}~\bibnamefont {Lin}}, \bibinfo {author}
		{\bibfnamefont {R.~R.~P.}\ \bibnamefont {Singh}}, \bibinfo {author} {\bibfnamefont {B.~D.}\ \bibnamefont {Gaulin}}, \ and\ \bibinfo {author}
		{\bibfnamefont {M.~J.~P.}\ \bibnamefont {Gingras}},\ }\href@noop {}
	{\bibfield  {journal} {\bibinfo  {journal} {Physical Review B}\ }\textbf
		{\bibinfo {volume} {87}},\ \bibinfo {pages} {184423} (\bibinfo {year}
		{2013})},\ \Eprint {http://arxiv.org/abs/1211.5934} {1211.5934} \BibitemShut
	{NoStop}%
	\bibitem [{\citenamefont {Ross}\ \emph {et~al.}(2012)\citenamefont {Ross},
		\citenamefont {Proffen}, \citenamefont {Dabkowska}, \citenamefont {Quilliam},
		\citenamefont {Yaraskavitch}, \citenamefont {Kycia},\ and\ \citenamefont
		{Gaulin}}]{ross2012lightly}%
	\BibitemOpen
	\bibfield  {author} {\bibinfo {author} {\bibfnamefont {K.~A.}\ \bibnamefont
			{Ross}}, \bibinfo {author} {\bibfnamefont {T.~h.}\ \bibnamefont {Proffen}},
		\bibinfo {author} {\bibfnamefont {H.~A.}\ \bibnamefont {Dabkowska}}, \bibinfo
		{author} {\bibfnamefont {J.~A.}\ \bibnamefont {Quilliam}}, \bibinfo {author}
		{\bibfnamefont {L.~R.}\ \bibnamefont {Yaraskavitch}}, \bibinfo {author}
		{\bibfnamefont {J.~B.}\ \bibnamefont {Kycia}}, \ and\ \bibinfo {author}
		{\bibfnamefont {B.~D.}\ \bibnamefont {Gaulin}},\ }\href@noop {} {\bibfield
		{journal} {\bibinfo  {journal} {Physical Review B}\ }\textbf {\bibinfo
			{volume} {86}},\ \bibinfo {pages} {174424} (\bibinfo {year}
		{2012})}\BibitemShut {NoStop}%
	\bibitem [{\citenamefont {Sala}\ \emph {et~al.}(2014)\citenamefont {Sala},
		\citenamefont {Gutmann}, \citenamefont {Prabhakaran}, \citenamefont
		{Pomaranski}, \citenamefont {Mitchelitis}, \citenamefont {Kycia},
		\citenamefont {Porter}, \citenamefont {Castelnovo},\ and\ \citenamefont
		{Goff}}]{sala2014vacancy}%
	\BibitemOpen
	\bibfield  {author} {\bibinfo {author} {\bibfnamefont {G.}~\bibnamefont
			{Sala}}, \bibinfo {author} {\bibfnamefont {M.~J.}\ \bibnamefont {Gutmann}},
		\bibinfo {author} {\bibfnamefont {D.}~\bibnamefont {Prabhakaran}}, \bibinfo
		{author} {\bibfnamefont {D.}~\bibnamefont {Pomaranski}}, \bibinfo {author}
		{\bibfnamefont {C.}~\bibnamefont {Mitchelitis}}, \bibinfo {author}
		{\bibfnamefont {J.~B.}\ \bibnamefont {Kycia}}, \bibinfo {author}
		{\bibfnamefont {D.~G.}\ \bibnamefont {Porter}}, \bibinfo {author}
		{\bibfnamefont {C.}~\bibnamefont {Castelnovo}}, \ and\ \bibinfo {author}
		{\bibfnamefont {J.~P.}\ \bibnamefont {Goff}},\ }\href@noop {} {\bibfield
		{journal} {\bibinfo  {journal} {Nature Materials}\ }\textbf {\bibinfo
			{volume} {13}},\ \bibinfo {pages} {488} (\bibinfo {year} {2014})}\BibitemShut
	{NoStop}%
	\bibitem [{\citenamefont {Arpino}\ \emph {et~al.}(2017)\citenamefont {Arpino},
		\citenamefont {Trump}, \citenamefont {Scheie}, \citenamefont {McQueen},\ and\
		\citenamefont {Koohpayeh}}]{arpino2017impact}%
	\BibitemOpen
	\bibfield  {author} {\bibinfo {author} {\bibfnamefont {K.~E.}\ \bibnamefont
			{Arpino}}, \bibinfo {author} {\bibfnamefont {B.~A.}\ \bibnamefont {Trump}},
		\bibinfo {author} {\bibfnamefont {A.~O.}\ \bibnamefont {Scheie}}, \bibinfo
		{author} {\bibfnamefont {T.~M.}\ \bibnamefont {McQueen}}, \ and\ \bibinfo
		{author} {\bibfnamefont {S.~M.}\ \bibnamefont {Koohpayeh}},\ }\href@noop {}
	{\bibfield  {journal} {\bibinfo  {journal} {Physical Review B}\ }\textbf
		{\bibinfo {volume} {95}},\ \bibinfo {pages} {094407} (\bibinfo {year}
		{2017})}\BibitemShut {NoStop}%
	\bibitem [{Note1()}]{Note1}%
	\BibitemOpen
	\bibinfo {note} {This collection of relatively large (compared to powder
		samples) crystals represents a large sampling of random orientations, but
		does not exactly correspond to a powder average. We did not pulverize the
		crystals in order to avoid strain broadening of the $g$-tensor.}\BibitemShut
	{Stop}%
	\bibitem [{\citenamefont {Van~Tol}\ \emph {et~al.}(2005)\citenamefont
		{Van~Tol}, \citenamefont {Brunel},\ and\ \citenamefont
		{Wylde}}]{van2005quasioptical}%
	\BibitemOpen
	\bibfield  {author} {\bibinfo {author} {\bibfnamefont {J.}~\bibnamefont
			{Van~Tol}}, \bibinfo {author} {\bibfnamefont {L.-C.}\ \bibnamefont {Brunel}},
		\ and\ \bibinfo {author} {\bibfnamefont {R.~J.}\ \bibnamefont {Wylde}},\
	}\href@noop {} {\bibfield  {journal} {\bibinfo  {journal} {Review of
				Scientific Instruments}\ }\textbf {\bibinfo {volume} {76}},\ \bibinfo {pages}
		{074101} (\bibinfo {year} {2005})}\BibitemShut {NoStop}%
	\bibitem [{\citenamefont {Rodriguez}\ \emph {et~al.}(2008)\citenamefont
		{Rodriguez}, \citenamefont {Adler}, \citenamefont {Brand}, \citenamefont
		{Broholm}, \citenamefont {Cook}, \citenamefont {Brocker}, \citenamefont
		{Hammond}, \citenamefont {Huang}, \citenamefont {Hundertmark}, \citenamefont
		{Lynn} \emph {et~al.}}]{rodriguez2008macs}%
	\BibitemOpen
	\bibfield  {author} {\bibinfo {author} {\bibfnamefont {J.~A.}\ \bibnamefont
			{Rodriguez}}, \bibinfo {author} {\bibfnamefont {D.~M.}\ \bibnamefont
			{Adler}}, \bibinfo {author} {\bibfnamefont {P.~C.}\ \bibnamefont {Brand}},
		\bibinfo {author} {\bibfnamefont {C.}~\bibnamefont {Broholm}}, \bibinfo
		{author} {\bibfnamefont {J.~C.}\ \bibnamefont {Cook}}, \bibinfo {author}
		{\bibfnamefont {C.}~\bibnamefont {Brocker}}, \bibinfo {author} {\bibfnamefont
			{R.}~\bibnamefont {Hammond}}, \bibinfo {author} {\bibfnamefont
			{Z.}~\bibnamefont {Huang}}, \bibinfo {author} {\bibfnamefont
			{P.}~\bibnamefont {Hundertmark}}, \bibinfo {author} {\bibfnamefont {J.~W.}\
			\bibnamefont {Lynn}},  \emph {et~al.},\ }\href@noop {} {\bibfield  {journal}
		{\bibinfo  {journal} {Measurement Science and Technology}\ }\textbf {\bibinfo
			{volume} {19}},\ \bibinfo {pages} {034023} (\bibinfo {year}
		{2008})}\BibitemShut {NoStop}%
	\bibitem [{\citenamefont {Ehlers}\ \emph {et~al.}(2011)\citenamefont {Ehlers},
		\citenamefont {Podlesnyak}, \citenamefont {Niedziela}, \citenamefont
		{Iverson},\ and\ \citenamefont {Sokol}}]{ehlers2011new}%
	\BibitemOpen
	\bibfield  {author} {\bibinfo {author} {\bibfnamefont {G.}~\bibnamefont
			{Ehlers}}, \bibinfo {author} {\bibfnamefont {A.~A.}\ \bibnamefont
			{Podlesnyak}}, \bibinfo {author} {\bibfnamefont {J.~L.}\ \bibnamefont
			{Niedziela}}, \bibinfo {author} {\bibfnamefont {E.~B.}\ \bibnamefont
			{Iverson}}, \ and\ \bibinfo {author} {\bibfnamefont {P.~E.}\ \bibnamefont
			{Sokol}},\ }\href@noop {} {\bibfield  {journal} {\bibinfo  {journal} {Review
				of Scientific Instruments}\ }\textbf {\bibinfo {volume} {82}},\ \bibinfo
		{pages} {085108} (\bibinfo {year} {2011})}\BibitemShut {NoStop}%
	\bibitem [{\citenamefont {Tang}\ \emph
		{et~al.}(2013{\natexlab{a}})\citenamefont {Tang}, \citenamefont {Khatami},\
		and\ \citenamefont {Rigol}}]{tang2013short}%
	\BibitemOpen
	\bibfield  {author} {\bibinfo {author} {\bibfnamefont {B.}~\bibnamefont
			{Tang}}, \bibinfo {author} {\bibfnamefont {E.}~\bibnamefont {Khatami}}, \
		and\ \bibinfo {author} {\bibfnamefont {M.}~\bibnamefont {Rigol}},\
	}\href@noop {} {\bibfield  {journal} {\bibinfo  {journal} {Computer Physics
				Communications}\ }\textbf {\bibinfo {volume} {184}},\ \bibinfo {pages} {557}
		(\bibinfo {year} {2013}{\natexlab{a}})}\BibitemShut {NoStop}%
	\bibitem [{\citenamefont {Jaubert}\ \emph {et~al.}(2015)\citenamefont
		{Jaubert}, \citenamefont {Benton}, \citenamefont {Rau}, \citenamefont
		{Oitmaa}, \citenamefont {Singh}, \citenamefont {Shannon},\ and\ \citenamefont
		{Gingras}}]{PhysRevLett.115.267208}%
	\BibitemOpen
	\bibfield  {author} {\bibinfo {author} {\bibfnamefont {L.~D.~C.}\
			\bibnamefont {Jaubert}}, \bibinfo {author} {\bibfnamefont {O.}~\bibnamefont
			{Benton}}, \bibinfo {author} {\bibfnamefont {J.~G.}\ \bibnamefont {Rau}},
		\bibinfo {author} {\bibfnamefont {J.}~\bibnamefont {Oitmaa}}, \bibinfo
		{author} {\bibfnamefont {R.~R.~P.}\ \bibnamefont {Singh}}, \bibinfo {author}
		{\bibfnamefont {N.}~\bibnamefont {Shannon}}, \ and\ \bibinfo {author}
		{\bibfnamefont {M.~J.~P.}\ \bibnamefont {Gingras}},\ }\href@noop {}
	{\bibfield  {journal} {\bibinfo  {journal} {Physical Review Letters}\
		}\textbf {\bibinfo {volume} {115}},\ \bibinfo {pages} {267208} (\bibinfo
		{year} {2015})}\BibitemShut {NoStop}%
	\bibitem [{\citenamefont {Elhajal}\ \emph {et~al.}(2005)\citenamefont
		{Elhajal}, \citenamefont {Canals}, \citenamefont {Sunyer},\ and\
		\citenamefont {Lacroix}}]{canalsdm1}%
	\BibitemOpen
	\bibfield  {author} {\bibinfo {author} {\bibfnamefont {M.}~\bibnamefont
			{Elhajal}}, \bibinfo {author} {\bibfnamefont {B.}~\bibnamefont {Canals}},
		\bibinfo {author} {\bibfnamefont {R.}~\bibnamefont {Sunyer}}, \ and\ \bibinfo
		{author} {\bibfnamefont {C.}~\bibnamefont {Lacroix}},\ }\href@noop {}
	{\bibfield  {journal} {\bibinfo  {journal} {Physical Review B}\ }\textbf
		{\bibinfo {volume} {71}},\ \bibinfo {pages} {094420} (\bibinfo {year}
		{2005})}\BibitemShut {NoStop}%
	\bibitem [{\citenamefont {Canals}\ \emph {et~al.}(2008)\citenamefont {Canals},
		\citenamefont {Elhajal},\ and\ \citenamefont {Lacroix}}]{canalsdm2}%
	\BibitemOpen
	\bibfield  {author} {\bibinfo {author} {\bibfnamefont {B.}~\bibnamefont
			{Canals}}, \bibinfo {author} {\bibfnamefont {M.}~\bibnamefont {Elhajal}}, \
		and\ \bibinfo {author} {\bibfnamefont {C.}~\bibnamefont {Lacroix}},\
	}\href@noop {} {\bibfield  {journal} {\bibinfo  {journal} {Physical Review
				B}\ }\textbf {\bibinfo {volume} {78}},\ \bibinfo {pages} {214431} (\bibinfo
		{year} {2008})}\BibitemShut {NoStop}%
	\bibitem [{\citenamefont {Chern}(2010)}]{chern2010pyrochlore}%
	\BibitemOpen
	\bibfield  {author} {\bibinfo {author} {\bibfnamefont {G.-W.}\ \bibnamefont
			{Chern}},\ }\href@noop {} {\bibfield  {journal} {\bibinfo  {journal} {arXiv
				preprint arXiv:1008.3038}\ } (\bibinfo {year} {2010})}\BibitemShut {NoStop}%
	\bibitem [{\citenamefont {Harris}\ \emph {et~al.}(1997)\citenamefont {Harris},
		\citenamefont {Bramwell}, \citenamefont {McMorrow}, \citenamefont {Zeiske},\
		and\ \citenamefont {Godfrey}}]{harris1997geometrical}%
	\BibitemOpen
	\bibfield  {author} {\bibinfo {author} {\bibfnamefont {M.~J.}\ \bibnamefont
			{Harris}}, \bibinfo {author} {\bibfnamefont {S.~T.}\ \bibnamefont
			{Bramwell}}, \bibinfo {author} {\bibfnamefont {D.~F.}\ \bibnamefont
			{McMorrow}}, \bibinfo {author} {\bibfnamefont {T.~H.}\ \bibnamefont
			{Zeiske}}, \ and\ \bibinfo {author} {\bibfnamefont {K.~W.}\ \bibnamefont
			{Godfrey}},\ }\href@noop {} {\bibfield  {journal} {\bibinfo  {journal}
			{Physical Review Letters}\ }\textbf {\bibinfo {volume} {79}},\ \bibinfo
		{pages} {2554} (\bibinfo {year} {1997})}\BibitemShut {NoStop}%
	\bibitem [{\citenamefont {Bramwell}\ and\ \citenamefont
		{Gingras}(2001)}]{bramwell2001spin}%
	\BibitemOpen
	\bibfield  {author} {\bibinfo {author} {\bibfnamefont {S.~T.}\ \bibnamefont
			{Bramwell}}\ and\ \bibinfo {author} {\bibfnamefont {M.~J.~P.}\ \bibnamefont
			{Gingras}},\ }\href@noop {} {\bibfield  {journal} {\bibinfo  {journal}
			{Science}\ }\textbf {\bibinfo {volume} {294}},\ \bibinfo {pages} {1495}
		(\bibinfo {year} {2001})}\BibitemShut {NoStop}%
	\bibitem [{\citenamefont {Benton}\ \emph {et~al.}(2016)\citenamefont {Benton},
		\citenamefont {Jaubert}, \citenamefont {Yan},\ and\ \citenamefont
		{Shannon}}]{benton2016spin}%
	\BibitemOpen
	\bibfield  {author} {\bibinfo {author} {\bibfnamefont {O.}~\bibnamefont
			{Benton}}, \bibinfo {author} {\bibfnamefont {L.~D.~C.}\ \bibnamefont
			{Jaubert}}, \bibinfo {author} {\bibfnamefont {H.}~\bibnamefont {Yan}}, \ and\
		\bibinfo {author} {\bibfnamefont {N.}~\bibnamefont {Shannon}},\ }\href@noop
	{} {\bibfield  {journal} {\bibinfo  {journal} {Nature Communications}\
		}\textbf {\bibinfo {volume} {7}},\ \bibinfo {pages} {11572} (\bibinfo {year}
		{2016})}\BibitemShut {NoStop}%
	\bibitem [{\citenamefont {Hermele}\ \emph {et~al.}(2004)\citenamefont
		{Hermele}, \citenamefont {Fisher},\ and\ \citenamefont
		{Balents}}]{hermele2004pyrochlore}%
	\BibitemOpen
	\bibfield  {author} {\bibinfo {author} {\bibfnamefont {M.}~\bibnamefont
			{Hermele}}, \bibinfo {author} {\bibfnamefont {M.~P.~A.}\ \bibnamefont
			{Fisher}}, \ and\ \bibinfo {author} {\bibfnamefont {L.}~\bibnamefont
			{Balents}},\ }\href@noop {} {\bibfield  {journal} {\bibinfo  {journal}
			{Physical Review B}\ }\textbf {\bibinfo {volume} {69}},\ \bibinfo {pages}
		{064404} (\bibinfo {year} {2004})}\BibitemShut {NoStop}%
	\bibitem [{\citenamefont {Molavian}\ \emph {et~al.}(2007)\citenamefont
		{Molavian}, \citenamefont {Gingras},\ and\ \citenamefont
		{Canals}}]{molavian2007}%
	\BibitemOpen
	\bibfield  {author} {\bibinfo {author} {\bibfnamefont {H.~R.}\ \bibnamefont
			{Molavian}}, \bibinfo {author} {\bibfnamefont {M.~J.~P.}\ \bibnamefont
			{Gingras}}, \ and\ \bibinfo {author} {\bibfnamefont {B.}~\bibnamefont
			{Canals}},\ }\href@noop {} {\bibfield  {journal} {\bibinfo  {journal}
			{Physical Review Letters}\ }\textbf {\bibinfo {volume} {98}},\ \bibinfo
		{pages} {157204} (\bibinfo {year} {2007})}\BibitemShut {NoStop}%
	\bibitem [{\citenamefont {Onoda}\ and\ \citenamefont
		{Tanaka}(2010)}]{onoda2009}%
	\BibitemOpen
	\bibfield  {author} {\bibinfo {author} {\bibfnamefont {S.}~\bibnamefont
			{Onoda}}\ and\ \bibinfo {author} {\bibfnamefont {Y.}~\bibnamefont {Tanaka}},\
	}\href@noop {} {\bibfield  {journal} {\bibinfo  {journal} {Physical Review
				Letters}\ }\textbf {\bibinfo {volume} {105}},\ \bibinfo {pages} {047201}
		(\bibinfo {year} {2010})}\BibitemShut {NoStop}%
	\bibitem [{\citenamefont {Gingras}\ and\ \citenamefont
		{McClarty}(2014)}]{gingras2014quantum}%
	\BibitemOpen
	\bibfield  {author} {\bibinfo {author} {\bibfnamefont {M.~J.~P.}\
			\bibnamefont {Gingras}}\ and\ \bibinfo {author} {\bibfnamefont {P.~A.}\
			\bibnamefont {McClarty}},\ }\href@noop {} {\bibfield  {journal} {\bibinfo
			{journal} {Reports on Progress in Physics}\ }\textbf {\bibinfo {volume}
			{77}},\ \bibinfo {pages} {056501} (\bibinfo {year} {2014})}\BibitemShut
	{NoStop}%
	\bibitem [{\citenamefont {Liu}\ \emph {et~al.}(2019)\citenamefont {Liu},
		\citenamefont {Hal{\'a}sz},\ and\ \citenamefont
		{Balents}}]{liu2019competing}%
	\BibitemOpen
	\bibfield  {author} {\bibinfo {author} {\bibfnamefont {C.}~\bibnamefont
			{Liu}}, \bibinfo {author} {\bibfnamefont {G.~B.}\ \bibnamefont {Hal{\'a}sz}},
		\ and\ \bibinfo {author} {\bibfnamefont {L.}~\bibnamefont {Balents}},\
	}\href@noop {} {\bibfield  {journal} {\bibinfo  {journal} {Physical Review
				B}\ }\textbf {\bibinfo {volume} {100}},\ \bibinfo {pages} {075125} (\bibinfo
		{year} {2019})}\BibitemShut {NoStop}%
	\bibitem [{\citenamefont {Banerjee}\ \emph {et~al.}(2016)\citenamefont
		{Banerjee}, \citenamefont {Bridges}, \citenamefont {Yan}, \citenamefont
		{Aczel}, \citenamefont {Li}, \citenamefont {Stone}, \citenamefont {Granroth},
		\citenamefont {Lumsden}, \citenamefont {Yiu}, \citenamefont {Knolle} \emph
		{et~al.}}]{banerjee2016proximate}%
	\BibitemOpen
	\bibfield  {author} {\bibinfo {author} {\bibfnamefont {A.}~\bibnamefont
			{Banerjee}}, \bibinfo {author} {\bibfnamefont {C.~A.}\ \bibnamefont
			{Bridges}}, \bibinfo {author} {\bibfnamefont {J.-Q.}\ \bibnamefont {Yan}},
		\bibinfo {author} {\bibfnamefont {A.~A.}\ \bibnamefont {Aczel}}, \bibinfo
		{author} {\bibfnamefont {L.}~\bibnamefont {Li}}, \bibinfo {author}
		{\bibfnamefont {M.~B.}\ \bibnamefont {Stone}}, \bibinfo {author}
		{\bibfnamefont {G.~E.}\ \bibnamefont {Granroth}}, \bibinfo {author}
		{\bibfnamefont {M.~D.}\ \bibnamefont {Lumsden}}, \bibinfo {author}
		{\bibfnamefont {Y.}~\bibnamefont {Yiu}}, \bibinfo {author} {\bibfnamefont
			{J.}~\bibnamefont {Knolle}},  \emph {et~al.},\ }\href@noop {} {\bibfield
		{journal} {\bibinfo  {journal} {Nature materials}\ }\textbf {\bibinfo
			{volume} {15}},\ \bibinfo {pages} {733} (\bibinfo {year} {2016})}\BibitemShut
	{NoStop}%
	\bibitem [{\citenamefont {Banerjee}\ \emph {et~al.}(2017)\citenamefont
		{Banerjee}, \citenamefont {Yan}, \citenamefont {Knolle}, \citenamefont
		{Bridges}, \citenamefont {Stone}, \citenamefont {Lumsden}, \citenamefont
		{Mandrus}, \citenamefont {Tennant}, \citenamefont {Moessner},\ and\
		\citenamefont {Nagler}}]{banerjee2017neutron}%
	\BibitemOpen
	\bibfield  {author} {\bibinfo {author} {\bibfnamefont {A.}~\bibnamefont
			{Banerjee}}, \bibinfo {author} {\bibfnamefont {J.}~\bibnamefont {Yan}},
		\bibinfo {author} {\bibfnamefont {J.}~\bibnamefont {Knolle}}, \bibinfo
		{author} {\bibfnamefont {C.~A.}\ \bibnamefont {Bridges}}, \bibinfo {author}
		{\bibfnamefont {M.~B.}\ \bibnamefont {Stone}}, \bibinfo {author}
		{\bibfnamefont {M.~D.}\ \bibnamefont {Lumsden}}, \bibinfo {author}
		{\bibfnamefont {D.~G.}\ \bibnamefont {Mandrus}}, \bibinfo {author}
		{\bibfnamefont {D.~A.}\ \bibnamefont {Tennant}}, \bibinfo {author}
		{\bibfnamefont {R.}~\bibnamefont {Moessner}}, \ and\ \bibinfo {author}
		{\bibfnamefont {S.~E.}\ \bibnamefont {Nagler}},\ }\href@noop {} {\bibfield
		{journal} {\bibinfo  {journal} {Science}\ }\textbf {\bibinfo {volume}
			{356}},\ \bibinfo {pages} {1055} (\bibinfo {year} {2017})}\BibitemShut
	{NoStop}%
	\bibitem [{\citenamefont {Kermarrec}\ \emph {et~al.}(2017)\citenamefont
		{Kermarrec}, \citenamefont {Gaudet}, \citenamefont {Fritsch}, \citenamefont
		{Khasanov}, \citenamefont {Guguchia}, \citenamefont {Ritter}, \citenamefont
		{Ross}, \citenamefont {Dabkowska},\ and\ \citenamefont
		{Gaulin}}]{kermarrec_NatComm}%
	\BibitemOpen
	\bibfield  {author} {\bibinfo {author} {\bibfnamefont {E.}~\bibnamefont
			{Kermarrec}}, \bibinfo {author} {\bibfnamefont {J.}~\bibnamefont {Gaudet}},
		\bibinfo {author} {\bibfnamefont {K.}~\bibnamefont {Fritsch}}, \bibinfo
		{author} {\bibfnamefont {R.}~\bibnamefont {Khasanov}}, \bibinfo {author}
		{\bibfnamefont {Z.}~\bibnamefont {Guguchia}}, \bibinfo {author}
		{\bibfnamefont {C.}~\bibnamefont {Ritter}}, \bibinfo {author} {\bibfnamefont
			{K.~A.}\ \bibnamefont {Ross}}, \bibinfo {author} {\bibfnamefont {H.~A.}\
			\bibnamefont {Dabkowska}}, \ and\ \bibinfo {author} {\bibfnamefont {B.~D.}\
			\bibnamefont {Gaulin}},\ }\href@noop {} {\bibfield  {journal} {\bibinfo
			{journal} {Nature Communications}\ }\textbf {\bibinfo {volume} {8}},\
		\bibinfo {pages} {14810} (\bibinfo {year} {2017})}\BibitemShut {NoStop}%
	\bibitem [{\citenamefont {Ross}\ \emph {et~al.}(2009)\citenamefont {Ross},
		\citenamefont {Ruff}, \citenamefont {Adams}, \citenamefont {Gardner},
		\citenamefont {Dabkowska}, \citenamefont {Qiu}, \citenamefont {Copley},\ and\
		\citenamefont {Gaulin}}]{ross_rods}%
	\BibitemOpen
	\bibfield  {author} {\bibinfo {author} {\bibfnamefont {K.~A.}\ \bibnamefont
			{Ross}}, \bibinfo {author} {\bibfnamefont {J.~P.~C.}\ \bibnamefont {Ruff}},
		\bibinfo {author} {\bibfnamefont {C.~P.}\ \bibnamefont {Adams}}, \bibinfo
		{author} {\bibfnamefont {J.~S.}\ \bibnamefont {Gardner}}, \bibinfo {author}
		{\bibfnamefont {H.~A.}\ \bibnamefont {Dabkowska}}, \bibinfo {author}
		{\bibfnamefont {Y.}~\bibnamefont {Qiu}}, \bibinfo {author} {\bibfnamefont
			{J.~R.~D.}\ \bibnamefont {Copley}}, \ and\ \bibinfo {author} {\bibfnamefont
			{B.~D.}\ \bibnamefont {Gaulin}},\ }\href@noop {} {\bibfield  {journal}
		{\bibinfo  {journal} {Physical Review Letters}\ }\textbf {\bibinfo {volume}
			{103}},\ \bibinfo {pages} {227202} (\bibinfo {year} {2009})}\BibitemShut
	{NoStop}%
	\bibitem [{\citenamefont {Thompson}\ \emph {et~al.}(2011)\citenamefont
		{Thompson}, \citenamefont {McClarty}, \citenamefont {R\o{}nnow},
		\citenamefont {Regnault}, \citenamefont {Sorge},\ and\ \citenamefont
		{Gingras}}]{thompson_rods}%
	\BibitemOpen
	\bibfield  {author} {\bibinfo {author} {\bibfnamefont {J.~D.}\ \bibnamefont
			{Thompson}}, \bibinfo {author} {\bibfnamefont {P.~A.}\ \bibnamefont
			{McClarty}}, \bibinfo {author} {\bibfnamefont {H.~M.}\ \bibnamefont
			{R\o{}nnow}}, \bibinfo {author} {\bibfnamefont {L.~P.}\ \bibnamefont
			{Regnault}}, \bibinfo {author} {\bibfnamefont {A.}~\bibnamefont {Sorge}}, \
		and\ \bibinfo {author} {\bibfnamefont {M.~J.~P.}\ \bibnamefont {Gingras}},\
	}\href@noop {} {\bibfield  {journal} {\bibinfo  {journal} {Physical Review
				Letters}\ }\textbf {\bibinfo {volume} {106}},\ \bibinfo {pages} {187202}
		(\bibinfo {year} {2011})}\BibitemShut {NoStop}%
	\bibitem [{\citenamefont {Tokiwa}\ \emph {et~al.}(2016)\citenamefont {Tokiwa},
		\citenamefont {Yamashita}, \citenamefont {Udagawa}, \citenamefont {Kittaka},
		\citenamefont {Sakakibara}, \citenamefont {Terazawa}, \citenamefont
		{Shimoyama}, \citenamefont {Terashima}, \citenamefont {Yasui}, \citenamefont
		{Shibauchi} \emph {et~al.}}]{tokiwa2016possible}%
	\BibitemOpen
	\bibfield  {author} {\bibinfo {author} {\bibfnamefont {Y.}~\bibnamefont
			{Tokiwa}}, \bibinfo {author} {\bibfnamefont {T.}~\bibnamefont {Yamashita}},
		\bibinfo {author} {\bibfnamefont {M.}~\bibnamefont {Udagawa}}, \bibinfo
		{author} {\bibfnamefont {S.}~\bibnamefont {Kittaka}}, \bibinfo {author}
		{\bibfnamefont {T.}~\bibnamefont {Sakakibara}}, \bibinfo {author}
		{\bibfnamefont {D.}~\bibnamefont {Terazawa}}, \bibinfo {author}
		{\bibfnamefont {Y.}~\bibnamefont {Shimoyama}}, \bibinfo {author}
		{\bibfnamefont {T.}~\bibnamefont {Terashima}}, \bibinfo {author}
		{\bibfnamefont {Y.}~\bibnamefont {Yasui}}, \bibinfo {author} {\bibfnamefont
			{T.}~\bibnamefont {Shibauchi}},  \emph {et~al.},\ }\href@noop {} {\bibfield
		{journal} {\bibinfo  {journal} {Nature Communications}\ }\textbf {\bibinfo
			{volume} {7}},\ \bibinfo {pages} {10807} (\bibinfo {year}
		{2016})}\BibitemShut {NoStop}%
	\bibitem [{\citenamefont {Pan}\ \emph {et~al.}(2016)\citenamefont {Pan},
		\citenamefont {Laurita}, \citenamefont {Ross}, \citenamefont {Gaulin},\ and\
		\citenamefont {Armitage}}]{pan2016measure}%
	\BibitemOpen
	\bibfield  {author} {\bibinfo {author} {\bibfnamefont {L.}~\bibnamefont
			{Pan}}, \bibinfo {author} {\bibfnamefont {N.~J.}\ \bibnamefont {Laurita}},
		\bibinfo {author} {\bibfnamefont {K.~A.}\ \bibnamefont {Ross}}, \bibinfo
		{author} {\bibfnamefont {B.~D.}\ \bibnamefont {Gaulin}}, \ and\ \bibinfo
		{author} {\bibfnamefont {N.~P.}\ \bibnamefont {Armitage}},\ }\href@noop {}
	{\bibfield  {journal} {\bibinfo  {journal} {Nature Physics}\ }\textbf
		{\bibinfo {volume} {12}},\ \bibinfo {pages} {361} (\bibinfo {year}
		{2016})}\BibitemShut {NoStop}%
	\bibitem [{MAC()}]{MACSsite}%
	\BibitemOpen
	\href@noop {} {\enquote {\bibinfo {title} {Macs ii overview},}\ }\bibinfo
	{howpublished} {\url{
			https://www.ncnr.nist.gov/instruments/macs/Overview.html}},\ \bibinfo {note}
	{accessed: 2019-09-20}\BibitemShut {NoStop}%
	\bibitem [{CNC({\natexlab{a}})}]{CNCSsite}%
	\BibitemOpen
	\href@noop {} {\enquote {\bibinfo {title} {Cold neutron chopper
				spectrometer},}\ }\bibinfo {howpublished} {\url{
			https://neutrons.ornl.gov/cncs}} ({\natexlab{a}}),\ \bibinfo {note}
	{accessed: 2020-02-20}\BibitemShut {NoStop}%
	\bibitem [{CNC({\natexlab{b}})}]{CNCSressite}%
	\BibitemOpen
	\href@noop {} {\enquote {\bibinfo {title} {Cncs resolution},}\ }\bibinfo
	{howpublished} {\url{ https://rez.mcvine.ornl.gov/}} ({\natexlab{b}}),\
	\bibinfo {note} {accessed: 2020-02-20}\BibitemShut {NoStop}%
	\bibitem [{\citenamefont {Bertin}\ \emph {et~al.}(2012)\citenamefont {Bertin},
		\citenamefont {Chapuis}, \citenamefont {de~R{\'e}otier},\ and\ \citenamefont
		{Yaouanc}}]{bertin2012crystal}%
	\BibitemOpen
	\bibfield  {author} {\bibinfo {author} {\bibfnamefont {A.}~\bibnamefont
			{Bertin}}, \bibinfo {author} {\bibfnamefont {Y.}~\bibnamefont {Chapuis}},
		\bibinfo {author} {\bibfnamefont {P.~D.}\ \bibnamefont {de~R{\'e}otier}}, \
		and\ \bibinfo {author} {\bibfnamefont {A.}~\bibnamefont {Yaouanc}},\
	}\href@noop {} {\bibfield  {journal} {\bibinfo  {journal} {Journal of
				Physics: Condensed Matter}\ }\textbf {\bibinfo {volume} {24}},\ \bibinfo
		{pages} {256003} (\bibinfo {year} {2012})}\BibitemShut {NoStop}%
	\bibitem [{\citenamefont {Gaudet}\ \emph {et~al.}(2015)\citenamefont {Gaudet},
		\citenamefont {Maharaj}, \citenamefont {Sala}, \citenamefont {Kermarrec},
		\citenamefont {Ross}, \citenamefont {Dabkowska}, \citenamefont {Kolesnikov},
		\citenamefont {Granroth},\ and\ \citenamefont {Gaulin}}]{gaudet2015cef}%
	\BibitemOpen
	\bibfield  {author} {\bibinfo {author} {\bibfnamefont {J.}~\bibnamefont
			{Gaudet}}, \bibinfo {author} {\bibfnamefont {D.~D.}\ \bibnamefont {Maharaj}},
		\bibinfo {author} {\bibfnamefont {G.}~\bibnamefont {Sala}}, \bibinfo {author}
		{\bibfnamefont {E.}~\bibnamefont {Kermarrec}}, \bibinfo {author}
		{\bibfnamefont {K.~A.}\ \bibnamefont {Ross}}, \bibinfo {author}
		{\bibfnamefont {H.~A.}\ \bibnamefont {Dabkowska}}, \bibinfo {author}
		{\bibfnamefont {A.~I.}\ \bibnamefont {Kolesnikov}}, \bibinfo {author}
		{\bibfnamefont {G.~E.}\ \bibnamefont {Granroth}}, \ and\ \bibinfo {author}
		{\bibfnamefont {B.~D.}\ \bibnamefont {Gaulin}},\ }\href {\doibase
		10.1103/PhysRevB.92.134420} {\bibfield  {journal} {\bibinfo  {journal}
			{Physical Review B}\ }\textbf {\bibinfo {volume} {92}},\ \bibinfo {pages}
		{134420} (\bibinfo {year} {2015})}\BibitemShut {NoStop}%
	\bibitem [{\citenamefont {Brown}(2006)}]{inttablesc}%
	\BibitemOpen
	\bibfield  {author} {\bibinfo {author} {\bibfnamefont {P.}~\bibnamefont
			{Brown}},\ }\href {\doibase 10.1107/97809553602060000594} {\emph {\bibinfo
			{title} {International Tables for Crystallography}}},\ Vol.~\bibinfo {volume}
	{C}\ (\bibinfo  {publisher} {Springer},\ \bibinfo {year} {2006})\ Chap.\
	\bibinfo {chapter} {4.4.5}, pp.\ \bibinfo {pages} {454--461}\BibitemShut
	{NoStop}%
	\bibitem [{\citenamefont {Rigol}\ \emph {et~al.}(2006)\citenamefont {Rigol},
		\citenamefont {Bryant},\ and\ \citenamefont {Singh}}]{nlc1}%
	\BibitemOpen
	\bibfield  {author} {\bibinfo {author} {\bibfnamefont {M.}~\bibnamefont
			{Rigol}}, \bibinfo {author} {\bibfnamefont {T.}~\bibnamefont {Bryant}}, \
		and\ \bibinfo {author} {\bibfnamefont {R.~R.~P.}\ \bibnamefont {Singh}},\
	}\href {\doibase 10.1103/PhysRevLett.97.187202} {\bibfield  {journal}
		{\bibinfo  {journal} {Physical Review Letters}\ }\textbf {\bibinfo {volume}
			{97}},\ \bibinfo {pages} {187202} (\bibinfo {year} {2006})}\BibitemShut
	{NoStop}%
	\bibitem [{\citenamefont {Rigol}\ \emph {et~al.}(2007)\citenamefont {Rigol},
		\citenamefont {Bryant},\ and\ \citenamefont {Singh}}]{nlc2}%
	\BibitemOpen
	\bibfield  {author} {\bibinfo {author} {\bibfnamefont {M.}~\bibnamefont
			{Rigol}}, \bibinfo {author} {\bibfnamefont {T.}~\bibnamefont {Bryant}}, \
		and\ \bibinfo {author} {\bibfnamefont {R.~R.~P.}\ \bibnamefont {Singh}},\
	}\href {\doibase 10.1103/PhysRevE.75.061118} {\bibfield  {journal} {\bibinfo
			{journal} {Physical Review E}\ }\textbf {\bibinfo {volume} {75}},\ \bibinfo
		{pages} {061118} (\bibinfo {year} {2007})}\BibitemShut {NoStop}%
\end{thebibliography}
\end{document}